\documentclass[%
reprint,
superscriptaddress,
 amsmath,amssymb,
 aps,
]{revtex4-1}

\usepackage{color}

\usepackage{graphicx}
\usepackage{dcolumn}
\usepackage{bm}
\usepackage{physics}%

\usepackage{hyperref}

\usepackage{braket}
\usepackage{mathtools}
\usepackage{comment}

\begin{document}
\preprint{APS/123-QED}

\title{Exact results for nonlinear Drude weights in the spin-1/2 XXZ chain}

\author{Yuhi Tanikawa}
\affiliation{%
Department of Physics, The University of Tokyo, 7-3-1 Hongo, Bunkyo-ku, Tokyo 113-0033, Japan}%

\author{Kazuaki Takasan}
\affiliation{%
Department of Physics, University of California, Berkeley, California 94720, USA}%
\affiliation{%
Materials Sciences Division, Lawrence Berkeley National Laboratory, Berkeley, CA 94720, USA}%

\author{Hosho Katsura}
\affiliation{%
Department of Physics, The University of Tokyo, 7-3-1 Hongo, Bunkyo-ku, Tokyo 113-0033, Japan}%
\affiliation{%
Institute for Physics of Intelligence, The University of Tokyo, 7-3-1 Hongo, Bunkyo-ku, Tokyo 113-0033, Japan}%
\affiliation{%
Trans-scale Quantum Science Institute, University of Tokyo, Bunkyo-ku, Tokyo 113-0033, Japan}%

\date{\today}
\begin{abstract}
Nonlinear Drude weight (NLDW) is a generalization of the linear Drude weight, which characterizes the nonlinear transport in quantum many-body systems. We investigate these weights for the spin-1/2 XXZ chain in the critical regime. The effects of the Dzyaloshinskii–Moriya interaction and an external magnetic field are also studied. Solving the Bethe equations numerically, we obtain these weights for very large system sizes and identify parameter regimes where the weights diverge in the thermodynamic limit. These divergences appear in all the orders studied in this paper and can be regarded as a generic feature of the NLDWs. We study the origin of these divergences and reveal that they result from \textit{nonanalytic} finite-size corrections to the ground state energy. Furthermore, we compute closed-form expressions for several weights in the thermodynamic limit and find excellent agreement with the numerical results.
\end{abstract}

\maketitle

\textit{Introduction. ---}
Transport phenomena have been one of the most important subjects in condensed matter and statistical physics. The linear transport phenomena are well explained by the famous linear response theory~\cite{Kubo} and widely applied to many experiments. On the other hand, the nonlinear responses are less understood~\cite{Shimizu1} and we still do not have a systematic understanding of them. While the nonlinear responses have been well-studied in the field of nonlinear optics~\cite{Boyd_book, Bloembergen_book}, they are still an intriguing topic. For instance, rectification currents~\cite{Tan2016, Tokura2018} and high-harmonic generations~\cite{Kruchinin2018, Ghimire2019} in solids are experimentally observed and extensively studied recently. They are used as new experimental probes and expected to be utilized for future optical/electric devises. Stimulated by this situation, the theoretical investigation for nonlinear responses is rapidly developing~\cite{Sodemann2015, Morimoto2016, deJuan2017, Dan2019, Isobe2020, Ahn2020, Takasan2020}, but further studies are still desired. In particular, the understanding of the nonlinear responses in many-body interacting systems is still poor compared with the non-interacting case~\cite{Morimoto2018, Avdoshkin2020, Michishita2021}.

Very recently, \textit{nonlinear Drude weights} (NLDWs) characterizing the nonlinear static transport have been introduced~\cite{Watanabe-Oshikawa, Watanabe-Oshikawa-Liu}. 
This is an extension of the linear Drude weight which was introduced by Kohn as an indicator to distinguish metals and insulators in quantum many-body systems~\cite{Kohn} and has been extensively studied in various contexts related to transport phenomena. In particular, the Drude weight is calculable with the exact solutions of one-dimensional quantum many-body systems and thus has been a principal quantity in the studies of their transport phenomena at zero and finite temperature~\cite{Bertini2020}. As the linear one has played a very important role, the NLDWs are also expected to provide useful information about nonlinear transport even in interacting many-body systems. However, most of the properties of NLDWs are still unexplored. For example, Ref.~\cite{Watanabe-Oshikawa} reported the divergent behavior of the third-order Drude weight in the spin-1/2 XXZ chain.  This is regarded as a feature of NLDWs not existing in linear Drude weights, and calls for a more detailed analysis of NLDWs, especially in interacting systems.

\begin{figure*}[tbp]
  \includegraphics[width=\hsize]{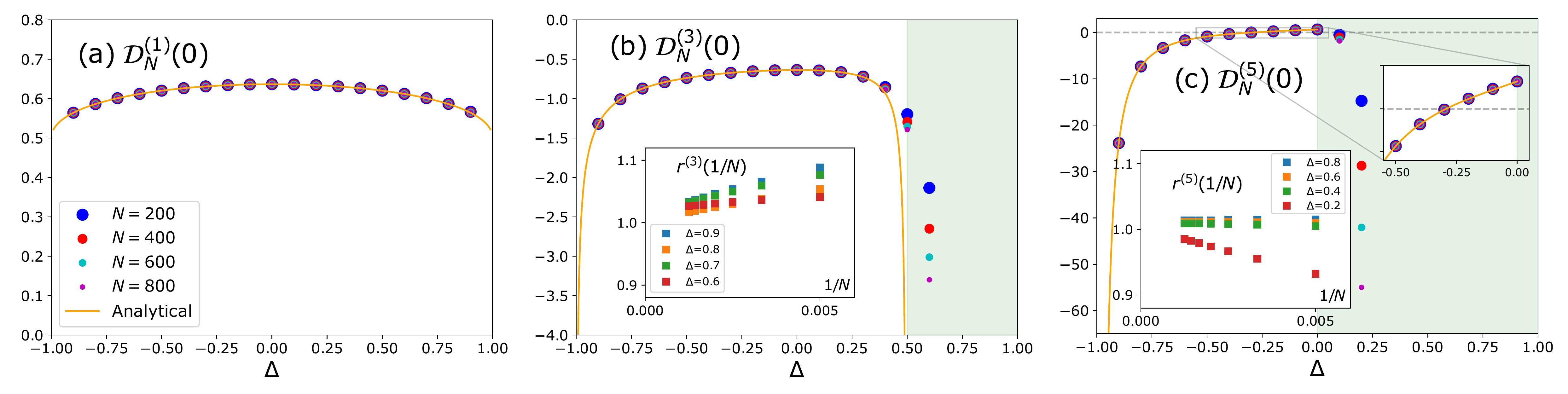}
  \vspace{-0.6cm}
  \caption{Numerical and analytical results for $\mathcal{D}_{N}^{(n)}(0)$ $(n=1,3,5)$. All the vertical axes are scaled with $J$.
  NLDWs $\mathcal{D}_{N}^{(3)}(\Theta)$ and $\mathcal{D}_{N}^{(5)}(\Theta)$ diverge in green regions, which are determined by $4\gamma/(\pi-\gamma)<n-1$. The insets in (b) and (c) show $r^{(n)}(1/N)$ [Eq.~(\ref{eq:r_odd})] in the divergent regions and confirm the divergence caused by {\it noninteger} powers of $1/N$.}
  \label{DW_of_XXZ}
\end{figure*}

In this paper, we study the NLDWs at zero temperature in the spin-1/2 XXZ chain, which is a prototypical many-body interacting model~\cite{FN1}. The linear Drude weight of this model has been  extensively studied in quantum transport phenomena~\cite{Sutherland, Shastry1990, Korepin1991, Narozhny1998, Zotos1999, Benz2005, Bertini2020}. The most important advantage of this model is its solvability by the Bethe ansatz~\cite{Takahashi, Korepin_book}, which enables us to treat very large system sizes. We also study the effect of the Dzyaloshinskii–Moriya (DM) interaction with a uniform DM vector along the z axis~\cite{Alcaraz1990} and an external magnetic field which are treatable within the Bethe ansatz technique. 
By using the exact solutions, we calculate the first several orders of the NLDWs numerically and find parameter regimes where the weights diverge in the thermodynamic limit.
While this divergence never appears in the linear one, it appears in all the NLDWs studied in this paper. Thus, we consider that the divergent behavior is one of the generic features of the NLDWs in interacting systems. To clarify the origin of this divergence, we analyze the finite size corrections to the ground state energy of the model. The detailed analysis shows that the divergence comes from a nonanalytic term proportional to a noninteger power of $1/N$ ($N$: system size). 
We explicitly identify the noninteger powers and confirmed  the  expected  divergence  by  using  our  numerical results.  Furthermore, we derive closed-form expressions
in the thermodynamic limit for several NLDWs in the convergent region by using the Wiener-Hopf method~\cite{Hamer, Takahashi, Sirker, Morse}. The obtained results match the numerical results with high accuracy.

\textit{Models. ---}
The spin-1/2 XXZ chain with periodic boundary conditions is defined by the Hamiltonian:
\begin{gather}
   \label{H_0}
   \hat{\mathcal{H}}(0)\!=\!\sum_{l=1}^{N}2J\bigg[\hat{S}_{l}^{x}\hat{S}_{l+1}^{x}+\hat{S}_{l}^{y}\hat{S}_{l+1}^{y}+\Delta \hat{S}_{l}^{z}\hat{S}_{l+1}^{z}\!\bigg],
\end{gather}
where $\hat{S}_{l}^{\alpha}$ $(\alpha=x,y,z)$ are spin-1/2 operators, $J>0$ is the coupling constant, $\Delta$ is the anisotropy parameter, and $N$ is the number of sites. We identify $N+1$ with $1$ and assume that $-1<\Delta<1$ and $N$ is even throughout this paper. Note that this model is mapped to the interacting spinless fermion model via the Jordan-Wigner transformation~\cite{Takahashi}. 
In this model, the Hamiltonian with the $U(1)$ flux $\Phi$ reads
  \begin{align}
    \label{H_DM}
    \hat{\mathcal{H}}(\Phi)
    &\!=\!\sum_{l=1}^{N}2J\bigg[\frac{1}{2}e^{i\frac{\Phi}{N}}\hat{S}_{l}^{+}\hat{S}_{l+1}^{-}\!\!+{\!\rm h.c.\!}+\!\Delta \hat{S}_{l}^{z}\hat{S}_{l+1}^{z}\!\bigg],\! 
  \end{align}
where $\hat{S}^\pm_l = \hat{S}^x_l \pm i \hat{S}^y_l$. Here it is enough to consider only $-\pi<\Phi\leq \pi$, as ${\cal H}(\Phi)$ and ${\cal H}(\Phi+2\pi)$ have the same spectrum. The $\Phi\neq0$ case corresponds to the spin-1/2 XXZ chain with the DM interaction~\cite{Alcaraz1990, FN2} When we consider the effect of an external magnetic field, we add to the Hamiltonian the term $-h\sum_{l=1}^{N}\hat{S}_{l}^{z}$ where $h$ is the magnetic field.

Since the total magnetization $\hat{S}^{z}_{\rm tot}=\sum_{l=1}^N\hat{S}^{z}_{l}$ is conserved even under the magnetic field, we can obtain the lowest energy state in each sector individually by the Bethe ansatz~\cite{Yang-Yang}. In the sector with $M$ down spins, the Bethe roots $\{v_j(\Phi)\}$ are determined by the following Bethe equation for $j=1,2,\ldots,M$:
  \begin{align}
    &p_1\big(v_j\qty(\Phi)\big)+\frac{\Phi}{N}-\frac{1}{N}\sum_{k=1}^{M}{p_2\big(v-v_k\qty(\Phi)\big)} \nonumber \\
    &\qquad \qquad \qquad \qquad \qquad=\frac{\pi}{N}\left(-M+2j-1\right), \label{eq:Bethe_eq}
  \end{align}
where $p_n(v)\equiv 2\tan^{-1}{\left(\frac{\tanh\frac{\gamma}{2}v}{\tan\frac{n\gamma}{2}}\right)}$ and $\gamma\equiv\arccos{\Delta}$. Using the Bethe roots, the energy density is given as 
\begin{align}
    \label{EGS}
    e(\Phi, h; M)
    &=\frac{1}{N}\sum_{j=1}^{M}\frac{2J\sin^2{\gamma}}{\cos{\gamma}-\cosh{\gamma {v_{j}(\Phi)}}}\nonumber\\
    &\qquad \qquad \quad+\frac{J\Delta}{2}-h\left(\frac{1}{2}-\frac{M}{N}\right).
\end{align}
If $h=0$ and $\Phi=0$, it is known that the ground state lies in the sector of $M=N/2$~\cite{Affleck}. Thus, for sufficiently small $\Phi$ the ground state energy density of $\mathcal{H}(\Phi)$ is $e_\mathrm{gs}(\Phi)=e(\Phi, h=0; M=N/2)$~\footnote{We have checked numerically for small system sizes that this relation holds for any $\Phi \in (-\pi, \pi)$}. Under the magnetic field $h$, $M$ is not necessarily equal to $N/2$ and the ground state energy density is given as $e_\mathrm{gs}(\Phi, h)=\min_M e(\Phi, h; M)$.

\begin{figure*}[tbp]
    \includegraphics[width=\hsize]{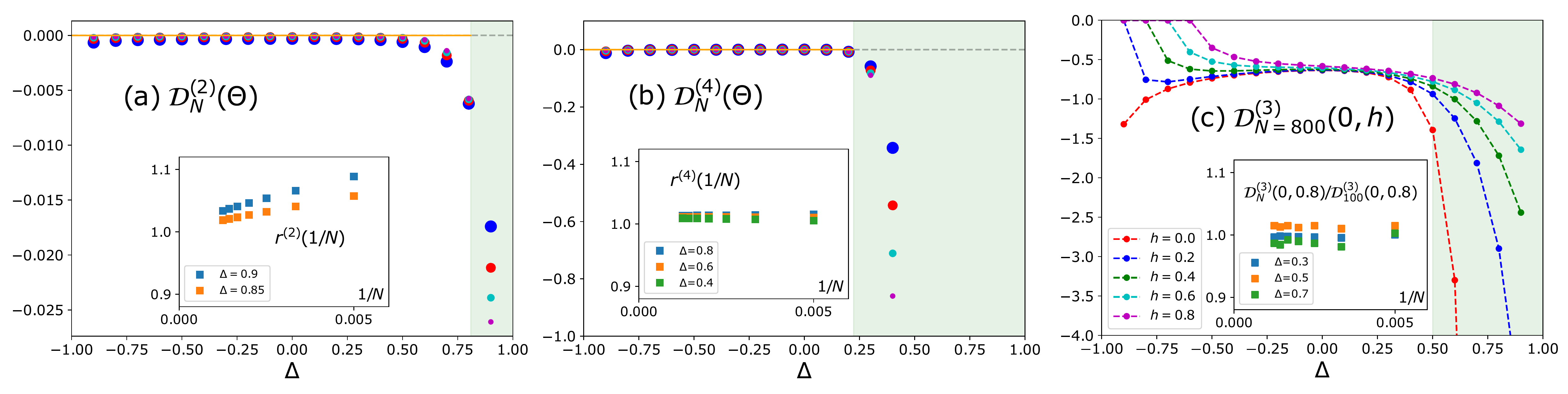}
    \vspace{-0.6cm}
    \caption{Numerical and analytical results for $\mathcal{D}_{N}^{(n)}(\Theta=0.1)$ $(n=2,4)$ are shown in (a) and (b). 
    Symbols are the same as in Fig.~(\ref{DW_of_XXZ}). 
    Numerical results for $\mathcal{D}_{N=800}^{(3)}(0,h)$ are shown in (c).
    All the vertical axes are scaled with $J$.
    Green regions are the divergent regions of NLDWs without a magnetic field, which are determined by $4\gamma/(\pi-\gamma)<n-1$. The insets in (a) and (b) show $r^{(n)}(1/N)$ [Eq.~(\ref{eq:r_even})] in the divergent regions and confirm the divergence caused by {\it noninteger} powers of $1/N$.
    The inset in (c) shows $\mathcal{D}_{N}^{(3)}(0,h=0.8)/\mathcal{D}_{100}^{(3)}(0,h=0.8)$ and confirms the convergence.}
    \label{DW_of_XXZ+DM}
\end{figure*}

\textit{Nonlinear Drude weight. ---}
Let us introduce the NLDWs. We follow the argument of Ref.~\cite{Watanabe-Oshikawa}. We consider the application of the time-dependent flux $\Phi(t)\equiv N\phi(t)$.
This induces the spin current density $j_{s}(t)=\langle \psi(t) |\partial \hat{\mathcal{H}}(\Phi) / \partial \Phi| \psi(t) \rangle$. Here, the state at time $t$ is defined as $\ket{\psi(t)}=\hat{U}(t)\ket{\psi_0}$ where  $\hat{U}(t)=\mathcal{T}\exp [-i\int^t_0 \hat{\mathcal{H}}(\Phi(s))ds]$ is the time-evolution operator and  $\ket{\psi_0}$ is  the ground state of $\hat{{\cal H}}(\Phi(0))$. Then, we define the linear and nonlinear conductivities in real time as
  \begin{align}
    \nonumber
    j_{s}(t)\!-\!j_{s}(0)
    \!&=\!\sum_{n=1}^{\infty}\frac{1}{n!}\int^{t}_{0}dt_{1}\cdots\int^{t}_{0}dt_{n}
    \\
    &\times\sigma^{(n)}(t-t_{1},\ldots,t-t_{n})\prod^{n}_{l=1}\!\qty(\!-\dv{\phi(t_{l})}{t_{l}}\!).
  \end{align}
Since the response function $\sigma^{(n)}(t_{1},\ldots,t_{n})$ vanishes whenever $t_{l}<0$ for any $l=1,2,\ldots,n$ due to the causality, the Fourier transform is given as
$\sigma^{(n)}(\omega_{1},\ldots,\omega_{n})
=\int^{\infty}_{0}dt_{1}\cdots\int^{\infty}_{0}dt_{n}\sigma^{(n)}(t_{1},\ldots,t_{n})\prod^{n}_{l=1}e^{i(\omega_{l}+i0)t_{l}}$.
The $n$-th order Drude weights in a finite system $\mathcal{D}^{(n)}_N$ are defined by the most singular part of $\sigma^{(n)}(\omega_{1},\ldots,\omega_{n})$ around $\omega_{1},\ldots,\omega_{n}=0$ and thus reads
\begin{align}
  \sigma^{(n)}_{\rm Drude}(\omega_{1},\ldots,\omega_{n})
    =\mathcal{D}^{(n)}_N\prod_{l=1}^{n} \frac{i}{\omega_{l}+i0}, \label{eq:def}
\end{align}
where the $n$-th order conductivity is decomposed as $\sigma^{(n)}=\sigma^{(n)}_\mathrm{Drude}+\sigma^{(n)}_\mathrm{regular}$~\cite{FN1}. At zero temperature, NLDWs $\mathcal{D}_N^{(n)}$ can be calculated as
\begin{gather}
    \mathcal{D}^{(n)}_N(\Theta)
    =N^{n+1}\pdv[n+1]{\Phi}e_{\rm gs}(\Phi)\Bigr|_{\Phi=\Theta}. \label{eq:Kohn_formula}
\end{gather}
This is the one-dimensional version of the nonlinear Kohn formula derived in Refs.~\cite{Watanabe-Oshikawa} and \cite{Watanabe-Oshikawa-Liu} which provide two different derivations, respectively. The finite $\Theta$ corresponds to the DM interaction as mentioned above. Under a finite magnetic field, we define $\mathcal{D}^{(n)}_N(\Theta, h)$ with replacing $e_\mathrm{gs}(\Phi)$  by $e_\mathrm{gs}(\Phi, h)$ in Eq.~(\ref{eq:Kohn_formula}). Note that the spin current corresponds to the electric (particle) current when the spin chain is mapped to the fermionic chain and thus the NLDWs defined above are related not only to the spin transport but also to more generic transport properties in interacting many-body systems.

\textit{Numerical results. ---}
By numerically solving the Bethe equations [Eq.~(\ref{eq:Bethe_eq})], we calculate the NLDWs $\mathcal{D}^{(n)}_N(\Theta)$. To calculate them, we approximate the derivative in Eq.~(\ref{eq:Kohn_formula}) by finite differences. 

First, we study the $\Theta=0$ case where only the odd orders are nonvanishing. This is because the ground state energy density $e_{\rm gs}(\Phi)$ is an even function of $\Phi$~\footnote{This can be seen by noting that $[\prod_{l=1}^{N}2\hat{S}_{l}^{x}]^{-1}\hat{\mathcal{H}}(\Phi)[\prod_{l=1}^{N}2\hat{S}_{l}^{x}]=\hat{\mathcal{H}}(-\Phi)$.}. It corresponds to the fact that the even order nonlinear responses vanish in inversion symmetric systems, which is well-known in nonlinear optics~\cite{Boyd_book, Bloembergen_book}. The results for $\mathcal{D}_{N}^{(n)}(0)$ ($n=1,3,5$) are shown in Figs.~\ref{DW_of_XXZ}~(a)-(c).
Fig.~\ref{DW_of_XXZ}~(a) is consistent with the previous work~\cite{Sutherland}, and Fig.~\ref{DW_of_XXZ}~(b) is also consistent with the recent numerical results for small system sizes~\cite{Watanabe-Oshikawa}.
The most significant difference between the linear and nonlinear ones is the existence of divergent regions.
The third-order one $\mathcal{D}^{(3)}_N(0)$ and the fifth-order one $\mathcal{D}^{(5)}_N(0)$ tend to diverge for $0.5\lesssim\Delta<1$ and $0\lesssim\Delta<1$, respectively.
Note that $\mathcal{D}^{(5)}_N(0)$ crosses zero at $\Delta\simeq-0.29$ and changes its sign when passing through the point as seen in Fig.~\ref{DW_of_XXZ}~(c).
This is a unique feature which does not appear in the lower orders and there might arise some special properties at this point. We also note in passing that a divergent behavior similar to that of $\mathcal{D}^{(3)}_N(0)$ was found for the fourth derivative of the ground state energy density with respect to the magnetization~\cite{Nomura}. 

Next, we consider the $\Theta \neq 0$ case. As we mentioned, this corresponds to the XXZ spin chain with finite DM interaction which breaks the inversion symmetry. Thus, even order responses are allowed.
The results for $\mathcal{D}_{N}^{(n)}(\Theta\ne 0)$ ($n=2,4$) are shown in Figs.~\ref{DW_of_XXZ+DM}~(a)~and~(b).
As we expected,  $\mathcal{D}_{N}^{(n)}(\Theta\ne 0)$ is nonzero in finite systems.
We can see the convergence of $\mathcal{D}_{N}^{(n)}(\Theta\ne 0)$ ($n=2,4$) to $0$ in a wide range of $\Delta$ in the thermodynamic limit.
The interesting point is that there also exist the divergent regions, as seen in Figs.~\ref{DW_of_XXZ}~(b)~and~(c).
The second-order one $\mathcal{D}_{N}^{(2)}(\Theta\ne 0)$ and the forth-order one $\mathcal{D}_{N}^{(4)}(\Theta\ne 0)$ tend to diverge for $0.81\lesssim\Delta<1$ and $0.22\lesssim\Delta<1$, respectively.
Since the effect of the flux $\Theta$ is rewritten as a twisted boundary condition, 
the ground state energy density is expected to be independent of $\Theta$ in the thermodynamic limit. Thus, it might seem that $\mathcal{D}_{N}^{(n)}(\Theta\ne 0)$ is zero.
However, since the Drude weights are differential coefficients before taking the thermodynamic limit, the divergence does not contradict the above statement. This reflects that the thermodynamic limit and the differentiation with respect to $\Phi$ are not interchangeable.

Finally, we study the effect of the magnetic field. The results for $\mathcal{D}_{N=800}^{(3)}(0,h)$ are shown in Fig.~\ref{DW_of_XXZ+DM}~(c).
For the $\Delta$ around both $-1$ and $1$, the values are suppressed. Some of the values around $\Delta=-1$ reach zero. It is natural because the gapped regime comes into $|\Delta|<1$ under the magnetic field~\cite{Takahashi}. The more nontrivial one is around $\Delta=1$. It seems that the divergent behavior is suppressed by the magnetic field. Indeed, the $N$ dependence shown in the insets of Fig.~\ref{DW_of_XXZ+DM}~(c) confirms that the convergent region becomes wider under the magnetic field. As we discuss later, this behavior can be understood from the low-energy effective field theory.
Note that these properties are seen in other orders of NLDWs as well~\footnote{For the data of the other order weights under the magnetic field, see Supplemental Material}. 

\textit{Origin and properties of the divergence. ---}
As Figs.~\ref{DW_of_XXZ}~and~\ref{DW_of_XXZ+DM} imply, the NLDWs  diverge in the certain regions by taking the thermodynamic limit. While the same behavior in the third-order one was reported based on numerical  diagonalization~\cite{Watanabe-Oshikawa}, the origin remains unclear. Here, we show that these behaviors are caused by the special terms included in the power series expansion of $e_{\rm gs}(\Phi)$.
The finite size corrections to the ground-state energy of the XXZ spin chain have been studied in great detail~\cite{deVega1985, Hamer1986, Alcaraz1987PRL, Alcaraz1987AoP, Woynarovich, Lukyanov}. Previous studies revealed that the corrections include both integer and noninteger powers of $1/N$, both of which can be accounted for by considering the low-energy effective field theory of the XXZ chain, i.e., the $c=1$ conformal field theory (CFT) perturbed by irrelevant operators. Although the effect of the flux has not been fully explored, it is natural to assume that the coefficient of each correction term can be Taylor expanded around $\Phi=0$. This, together with the fact that $e_{\rm gs} (\Phi)$ is an even function of $\Phi$, yields
  \begin{align}
    \label{expansion}
    \nonumber
    e_{\rm gs}(\Phi)-e_{\rm gs}(0)=\sum_{k\geq l\geq1}A_{k,l}&\qty(\frac{1}{N})^{2k}\Phi^{2l}
    \\
    +\sum_{k,l,m\geq1}B_{k,l,m}&\qty(\frac{1}{N})^{2k+4m\frac{\gamma}{\pi-\gamma}}\Phi^{2l},
  \end{align}
where $A_{k,l}$ and $B_{k,l,m}$ are coefficients depending on the parameter $\gamma$. Note that the coefficients $A_{1,1}$, $A_{2,2}$, and $B_{1,l,1}$ can be read off from Eq.~$(4.1)$ in Ref.~\cite{Lukyanov}, and at the free-fermion point ($\gamma=\pi/2$), all the coefficients can be easily computed explicitly~\cite{Watanabe-Oshikawa}. 
In the following, for simplicity, we restrict ourselves to the case where $\gamma$ is generic, i.e., none of the exponents in the second sum are integers. 
The noninteger exponent can be rewritten as $2k+4m \gamma/(\pi-\gamma) = 2k+4m (2K-1)$, where $K=(\pi/2)/(\pi-\gamma)$ is the Tomonaga-Luttinger parameter of the model~\cite{Sirker, Giamarchi_book,SIRKER_2012}. 
This reflects that the nonanalytic finite size corrections originate from irrelevant operators with noninteger scaling dimensions such as $4K$~\cite{Alcaraz1987PRL, Alcaraz1987AoP}. 

Any term $\Phi^{\alpha}/N^{\beta}$ $(\alpha>\beta)$ in Eq.~(\ref{expansion}) can contribute to divergences of NLDWs.
The straightforward calculation gives 
  \begin{align}
  \label{2n-1}
    &\mathcal{D}^{(2k-1)}_N(\Theta)=(2k)!\bigg[A_{k,k}+B_{1,k,1}N^{2k-2-\frac{4\gamma}{\pi-\gamma}}+\cdots\bigg],
    \\\nonumber
    &\mathcal{D}^{(2k)}_N(\Theta)=(2k+2)!A_{k+1,k+1}\frac{\Theta}{N}
    \\\label{2n}
    &\qquad\qquad\qquad\qquad\quad +C_{k}(\Theta)N^{2k-1-\frac{4\gamma}{\pi-\gamma}}+\cdots,
  \end{align}
where $C_{k}(\Theta)\equiv\sum_{l> k}(2l)!/(2l-2k-1)!B_{1,l,1}\Theta^{2l-2k-1}$.
The above expressions suggest that  $\mathcal{D}^{(2k-1)}_N(\Theta)$ and $\mathcal{D}^{(2k)}_N(\Theta)$ are likely to diverge when the exponent of the power of $N$ in each second term, which can be the leading term, is positive: $2k>2+4\gamma/(\pi-\gamma)$ and $2k>1+4\gamma/(\pi-\gamma)$, respectively~\footnote{Although we have excluded non-generic values of $\gamma$, the leading behavior of the finite size correction is the same as Eqs. (9) and (10) even for these $\gamma$ except for $\gamma = \pi l/(l+2)\   (l\in\mathbb{N})$. The analysis of the exceptional cases requires some additional care and will be discussed elsewhere.}.
This enabled us to determine the green regions in Figs.~\ref{DW_of_XXZ}~and~\ref{DW_of_XXZ+DM}.

Also, Eqs.~(\ref{2n-1}) and (\ref{2n}) imply that $\mathcal{D}^{(n)}_N(\Theta)$ shows the 
divergence caused by $N^{n-1-\frac{4\gamma}{\pi-\gamma}}$ in the divergent region. In order to confirm this, we define  $r^{(2k-1)}(1/N)$ and $r^{(2k)}(1/N)$ as
\begin{align}
 r^{(2k-1)}(1/N)&=
 \frac{\mathcal{D}^{(2k-1)}_N(0)}{(2k)!B_{1,k,1}N^{2k-2-4\gamma/(\pi-\gamma)}},\label{eq:r_odd}
 \\ r^{(2k)}(1/N)&=\frac{ \mathcal{D}^{(2k)}_N(\Theta)}{C_{k}(\Theta)N^{2k-1-4\gamma/(\pi-\gamma)}},\label{eq:r_even}
\end{align}
and plot Eq.~(\ref{eq:r_odd}) [Eq.~(\ref{eq:r_even})] in the insets of Figs.~\ref{DW_of_XXZ}~(b) and (c) [Figs.~\ref{DW_of_XXZ+DM}~(a) and (b)].
These figures clearly show that each data is on a straight line to the value near $1$ in the large $N$ region, indicating that the divergences are caused by the {\it noninteger} power terms of $N$ expected from the power series expansion~(\ref{expansion}). We stress that the numerical confirmation of these behaviors is quite challenging because it requires large system sizes, which are beyond the reach of other numerical methods such as exact diagonalization.

The suppression of the divergence under the magnetic field around $\Delta=1$, shown in Fig.~\ref{DW_of_XXZ+DM}~(c), is also explained by the expansion (\ref{expansion}). 
In the absence of the magnetic field, the umklapp scattering term with scaling dimension $4K=2\pi/(\pi-\gamma)$ is responsible for the nonanalytic finite-size corrections. However, in the presence of the magnetic field, the Fermi wave vectors become incommensurate with the lattice. 
As a result, the umklapp term oscillates in space and should be dropped in a renormalization group sense~\cite{Sirker2011,Sirker, Giamarchi_book}. Therefore, the effect of noninteger powers in Eq.~(\ref{expansion}) are expected to be small under the magnetic field and thus the divergence is suppressed as well.

\textit{Analytical form in the convergent region.---}
By using the expansion [Eq.~(\ref{expansion})], we can derive closed-form expressions for NLDWs in the thermodynamic limit. Taking the limit in Eq.~(\ref{expansion}) in the convergent region, we obtain the NLDWs $\mathcal{D}^{(2k)}=0$ and $\mathcal{D}^{(2k-1)}=(2k)!A_{k,k}$ where $\mathcal{D}^{(n)}\equiv \lim_{N\rightarrow\infty}\mathcal{D}^{(n)}_N(\Theta)$, and thus the problem is reduced to the calculation of $A_{k,k}$. These coefficients can be calculated using the Wiener-Hopf method, which is a mathematical technique to solve the Wiener-Hopf integral equations~\cite{Hamer, Takahashi, Sirker, Morse} (see Supplemental Material for more details). 
As a result, the first-order (linear) one is
 \begin{gather}
    \label{AD1}
    \mathcal{D}^{(1)}=\frac{\pi J \sin{\gamma}}{2\gamma(\pi-\gamma)},
  \end{gather}
for $0<\gamma<\pi$. This reproduces the previous result in Refs.~\cite{Sutherland, Shastry1990}.
The third-order and fifth-order ones are  given by
\begin{gather}
    \label{AD3}
    \mathcal{D}^{(3)}=-\frac{J\sin{\gamma}}{8\gamma(\pi-\gamma)}
    \left[
    \frac{\Gamma\big(\frac{3\pi}{2\gamma}\big){\Gamma\big(\frac{\pi-\gamma}{2\gamma}\big)}^3}{\Gamma\big(\frac{3(\pi-\gamma)}{2\gamma}\big){\Gamma\big(\frac{\pi}{2\gamma}\big)}^3}
    +
    \frac{3\pi\tan{\big(\frac{\pi^{2}}{2\gamma}\big)}}{\pi-\gamma}
    \right],
\end{gather}
for $\pi/3<\gamma<\pi$, and
  \begin{align}
    \nonumber
    \mathcal{D}^{(5)}=&\frac{3J\sin{\gamma}}{32\pi\gamma(\pi-\gamma)}
    \left[
    \frac{\Gamma\big(\frac{5\pi}{2\gamma}\big){\Gamma\big(\frac{\pi-\gamma}{2\gamma}\big)}^5}{\Gamma\big(\frac{5(\pi-\gamma)}{2\gamma}\big){\Gamma\big(\frac{\pi}{2\gamma}\big)}^5}
    \right.
    \\\nonumber
    &\left.
    -
    \frac{5}{3}\cdot\frac{\Gamma\big(\frac{3\pi}{2\gamma}\big)^2{\Gamma\big(\frac{\pi-\gamma}{2\gamma}\big)}^6}{\Gamma\big(\frac{3(\pi-\gamma)}{2\gamma}\big)^2{\Gamma\big(\frac{\pi}{2\gamma}\big)}^6}
    +
    \frac{15\pi^2\tan^2{\big(\frac{\pi^{2}}{2\gamma}\big)}}{(\pi-\gamma)^2}
    \right.
    \\\label{AD5}
    &
    \qquad \quad
    \left.+
    \frac{5\pi\tan{\big(\frac{\pi^{2}}{2\gamma}\big)}}{\pi-\gamma}\cdot\frac{\Gamma\big(\frac{3\pi}{2\gamma}\big){\Gamma\big(\frac{\pi-\gamma}{2\gamma}\big)}^3}{\Gamma\big(\frac{3(\pi-\gamma)}{2\gamma}\big){\Gamma\big(\frac{\pi}{2\gamma}\big)}^3}
    \right],
  \end{align}
for $\pi/2<\gamma<\pi$, respectively. We note that the result of $\mathcal{D}^{(3)}$ can also be read off from $A_{2,2}$ in Eq. (\ref{expansion}) and is consistent with the result of Ref.~\cite{Watanabe-Oshikawa}. These are plotted in Fig.~\ref{DW_of_XXZ}. Clearly, the analytical results match the numerical results with high accuracy.

\textit{Conclusion and Outlook.---}
In this study, we calculated the zero-temperature NLDWs 
of the spin-1/2 XXZ chain in the critical regime numerically for large system sizes, considering the effect of the DM interaction or the external magnetic field. The numerical results [Figs.~\ref{DW_of_XXZ}~and~\ref{DW_of_XXZ+DM}] revealed that all the NLDWs diverge in certain $\Delta$ regions by taking the thermodynamic limit.
Thus, we considered these divergences are a generic feature in interacting systems and investigated their mechanism. 
Based on the power series expansion [Eq. (8)], we identified the origin of the divergences as nonanalytic finite-size corrections to the ground state energy.
This expansion also allows us to identify the regions and strength of the divergences. We confirmed that they are in good agreement with the numerical data.
Furthermore by using the Wiener-Hopf method, we obtained the closed forms of several weights in the thermodynamic limit [Eqs.~(\ref{AD1})-(\ref{AD5})]. In the convergent regions, they matched the numerical results with high accuracy as seen in Fig.~\ref{DW_of_XXZ}.
Although in this paper we calculated the analytical expressions for NLDWs by treating the magnetization and the U(1) flux simultaneously (see Supplemental Material), we expect that a direct calculation for zero magnetization should be possible using another method involving nonlinear integral equations~\cite{Klumper_1991}. A thorough analysis of NLDWs based on such a sophisticated method would be an interesting future direction. 

Our results are a first systematic calculation of the NLDWs in interacting many-body systems for very large system sizes. We found that the divergent behavior generically appears and clarified the origin of the divergence. We believe that our results will help understanding the nonlinear transport in quantum many-body systems.

\begin{acknowledgments}
We thank Haruki Watanabe, Masaki Oshikawa, and Kiyohide Nomura for valuable discussions. K. T. was supported by the U.S. Department of Energy (DOE), Office of Science, Basic Energy Sciences (BES), under Contract No. AC02-05CH11231 within the Ultrafast Materials Science Program (KC2203). K. T. thanks the Japan Society for the Promotion of Science (JSPS) for an Overseas Research Fellowship. H. K. was supported in part by JSPS Grant-in-Aid for Scientific Research on Innovative Areas No. JP20H04630, JSPS KAKENHI Grant No. JP18K03445, and the Inamori Foundation. 
\end{acknowledgments}

\bibliographystyle{apsrev4-1}
\bibliography{cite}

\clearpage


\appendix
\renewcommand{\theequation}{S\arabic{equation}}
\setcounter{equation}{0}
\renewcommand{\thefigure}{S\arabic{figure}}
\setcounter{figure}{0}

\begin{widetext}
\begin{center}
\textbf{\large Supplemental Material: ``Exact results for nonlinear Drude weights in quantum spin chains"}
\end{center}

\section*{S1. Dzyaloshinskii–Moriya interaction}
\label{DM}
Here we consider the spin-1/2 XXZ chain with DM interaction.
The model is defined by the Hamiltonian:
  \begin{align}
    \hat{\mathcal{H}}(0)+\sum_{l=1}^{N}{2\bm{D}_{l}}\cdot\big(\hat{\bm{S}}_{l}\times\hat{\bm{S}}_{l+1}\big)
     =&\sum_{l=1}^{N}\bigg[(J+iD)\hat{S}_{l}^{+}\hat{S}_{l+1}^{-}+{\rm h.c.}+2J\Delta \hat{S}_{l}^{z}\hat{S}_{l+1}^{z}\bigg]
     \\
     =&\sum_{l=1}^{N}2J_{\rm D}\bigg[\frac{1}{2}e^{i\theta}\hat{S}_{l}^{+}\hat{S}_{l+1}^{-}+{\rm h.c.}+\Delta_{\rm D} \hat{S}_{l}^{z}\hat{S}_{l+1}^{z}\bigg],
  \end{align}
where we assumed that the DM vector $\bm{D}_{k}$ is uniform and along the $z$ axiz, 
namely $\bm{D}_{k}=D\hat{\bm{z}}$, and introduced new parameters: $\theta\equiv\arg\qty(J+iD)$, $J_{\rm D}\equiv\sqrt{J^{2}+D^{2}}$, and $\Delta_{\rm D}\equiv J\Delta/\sqrt{J^{2}+D^{2}}$.
Then we can define the unitary transformed one as $\hat{\mathcal{H}}_{\rm DM}$:
  \begin{gather}
    \label{deriv_DM}
    \hat{\mathcal{H}}_{\rm DM}
    =\sum_{l=1}^{N}2J_{\rm D}\bigg[\frac{1}{2}e^{i\frac{\Theta}{N}}\hat{S}_{l}^{+}\hat{S}_{l+1}^{-}+{\rm h.c.}+\Delta_{\rm D} \hat{S}_{l}^{z}\hat{S}_{l+1}^{z}\bigg],
\end{gather}
where $\Theta$ is a uniquely determined constant satisfying $-\pi<\Theta\leq \pi$.
Thus renormalization enables us to identify the above with Eq.~(\ref{H_DM}).

\section*{S2. Numerical calculation of Drude weights}
In order to calculate $\mathcal{D}^{(n)}_N(\Theta)\ (n=1,2,3,4,5)$, we approximate the derivative in Eq.~(\ref{eq:Kohn_formula}) by finite differences as
  \begin{align}
    \mathcal{D}^{(1)}_N(\Theta)&\simeq N^2\frac{e_{\rm gs}(\delta+\Theta)+e_{\rm gs}(-\delta+\Theta)-2e_{\rm gs}(\Theta)}{\delta^2},
    \\
    \mathcal{D}^{(2)}_N(\Theta)&\simeq N^3\frac{e_{\rm gs}\qty(\frac{3\delta}{2}+\Theta)-e_{\rm gs}\qty(-\frac{3\delta}{2}+\Theta)-3\Big(e_{\rm gs}\qty(\frac{\delta}{2}+\Theta)-e_{\rm gs}\qty(-\frac{\delta}{2}+\Theta)\Big)}{\delta^3},
    \\
    \mathcal{D}^{(3)}_N(\Theta)&\simeq N^4\frac{e_{\rm gs}\qty(2\delta+\Theta)+e_{\rm gs}\qty(-2\delta+\Theta)-4\Big(e_{\rm gs}\qty(\delta+\Theta)+e_{\rm gs}\qty(-\delta+\Theta)\Big)+6e_{\rm gs}\qty(\Theta)}{\delta^4},
    \\
    \mathcal{D}^{(4)}_N(\Theta)&\simeq N^5\frac{e_{\rm gs}\qty(\frac{5\delta}{2}+\Theta)\!-\!e_{\rm gs}\qty(-\frac{5\delta}{2}+\Theta)\!-\!5\Big(e_{\rm gs}\qty(\frac{3\delta}{2}+\Theta)\!-\!e_{\rm gs}\qty(-\frac{3\delta}{2}+\Theta)\Big)\!+\!10\Big(e_{\rm gs}\qty(\frac{\delta}{2}+\Theta)\!-\!e_{\rm gs}\qty(-\frac{\delta}{2}+\Theta)\Big)}{\delta^5},
    \\
    \mathcal{D}^{(5)}_N(\Theta)&\simeq N^6\frac{e_{\rm gs}\qty(3\delta+\Theta)\!+\!e_{\rm gs}\qty(-3\delta+\Theta)\!-\!6\Big(e_{\rm gs}\qty(2\delta+\Theta)\!+\!e_{\rm gs}\qty(-2\delta+\Theta)\Big)\!+\!15\Big(e_{\rm gs}\qty(\delta+\Theta)\!+\!e_{\rm gs}\qty(-\delta+\Theta)\Big)\!-\!20e_{\rm gs}\qty(\Theta)}{\delta^6},
  \end{align}
where $\delta$ is sufficiently small. Note that too small $\delta$ may lead to numerical precision errors.
In order to calculate $\mathcal{D}^{(n)}_N(\Theta, h)$, we have to replace all the $e_\mathrm{gs}(\Phi)$ in the above relations by $e_\mathrm{gs}(\Phi, h)$.

\section*{S3. Wiener-Hopf method}
\label{Appendix_WH}
In this part, we introduce the Wiener-Hopf method {\cite{Hamer,Takahashi,Sirker,Morse}}.
The combination of this method and the Bethe ansatz enables us to calculate the lowest energy density of each magnetization sector in the thermodynamic limit.

\subsection{Bethe ansatz in the thermodynamic limit}
We consider the spin-1/2 XXZ chain without the magnetic field:
  \begin{gather}
    \hat{\mathcal{H}}(0)=\sum_{l=1}^{N}2J\bigg[\hat{S}_{l}^{x}\hat{S}_{l+1}^{x}+\hat{S}_{l}^{y}\hat{S}_{l+1}^{y}+\Delta \hat{S}_{l}^{z}\hat{S}_{l+1}^{z}\bigg].
  \end{gather}
Since the total magnetization $\hat{S}^{z}_{\rm tot}=\sum_{l}\hat{S}^{z}_{l}$ is conserved in this model, we can obtain the lowest energy state in each magnetization sector individually by the Bethe ansatz.
In the sector with $M$ down spins, the Bethe roots are determined by the following Bethe equations:
  \begin{gather}
    \mathcal{Z}_N(v_j)=\frac{2\pi I_j}{N}=\frac{\pi}{N}\left(-M+2j-1\right)
    \quad \left(j=1,2,\cdots,M \right),
  \end{gather}
where
  \begin{gather}
    \label{Z_N}
    \mathcal{Z}_N(v)\equiv p_1(v)-\frac{1}{N}\sum_{k=1}^{M}{p_2(v-v_k)}
  \end{gather}
with
  \begin{gather}
    p_n(v)\equiv 2\tan^{-1}{\left(\frac{\tanh\frac{\gamma}{2}v}{\tan\frac{n\gamma}{2}}\right)}.
  \end{gather}
It is known that there exists a unique set of real solutions $\{v_j\}$ satisfying $-\infty\leq v_1<v_2<\ldots<v_M\leq\infty$.
Differentiating Eq.~(\ref{Z_N}) with respect to $v$, we get
  \begin{gather}
    \label{rho_N}
    \rho_N(v)=a_1(v)-\frac{1}{N}\sum_{k=1}^{M}{a_2(v-v_k)},
  \end{gather}
where
  \begin{gather}
    \label{rho_N2}
    \rho_N(v)\equiv\frac{1}{2\pi}\dv{\mathcal{Z}_N(v)}{v}
  \end{gather}
and
  \begin{gather}
    a_n(v)\equiv\frac{1}{2\pi}\dv{v}p_n(v)=\frac{\gamma}{2\pi}\frac{\sin n\gamma}{\cosh\gamma v-\cos n\gamma}.
  \end{gather}
Then $\{v_j\}$ gives the lowest energy density as
  \begin{gather}
    \label{e_0}
    e(M)=-\frac{2\pi A}{N}\sum_{j=1}^{M}a_1(v_j)+\frac{\Delta}{2}, 
  \end{gather}
where $A=2J \sin \gamma/\gamma$. In the thermodynamic limit, the following relation holds:
  \begin{gather}
    \frac{1}{N}\sum_{j=1}^{M}f(v_j)=\int^{v_M}_{v_1}f(v)\rho_N(v)dv+\mathcal{O}\qty(N^{-1})
    \xrightarrow{N\rightarrow\infty}
    \int^{Q}_{-Q}f(v)\rho(v)dv,
  \end{gather}
where $f(v)$ is an arbitrary function of $\mathcal{O}(1)$ and $-Q,Q,\rho(v)$ are new representations of $v_1,v_M,\rho_N(v)$ in the limit, respectively.
Thus, Eqs.~(\ref{rho_N}) and (\ref{e_0}) can be expressed in the limit as follows:
  \begin{gather}
    \label{rho}
    \rho(v)=a_1(v)-\int^{Q}_{-Q}a_2(v-x)\rho(x)dx,
    \\
    \label{e0}
    e\qty(m)=-2\pi A\int^{Q}_{-Q}a_1(x)\rho(x)dx+\frac{\Delta}{2},
  \end{gather}
where $e\qty(m)$ is the new representation of $e(M)$.
Here we defined a new parameter $m$ as
  \begin{gather}
    \label{m}
    m\equiv\lim_{N\rightarrow\infty}\frac{N/2-M}{N}=\frac{1}{2}-\int^{Q}_{-Q}\rho(x)dx,
  \end{gather}
which corresponds to the magnetization.

\smallskip

Now we calculate the following value by using Eq.~(\ref{Z_N}):
  \begin{align}
    \nonumber
    \mathcal{Z}_N\big(\infty\big)-\mathcal{Z}_N\big(v_M\big)
    =&\qty(\pi-\gamma-\big(\pi-2\gamma\big)\frac{M}{N})-\frac{\pi}{N}\big(M-1\big)
    \\
    =&\frac{\pi}{N}+2\qty(\pi-\gamma)\qty(\frac{1}{2}-\frac{M}{N}).
  \end{align}
It follows from Eq.~(\ref{rho_N2}) that the left-hand side of the above equation can be expressed as
  \begin{gather}
    \mathcal{Z}_N\big(\infty\big)-\mathcal{Z}_N\big(v_M\big)
    =2\pi\int^{\infty}_{v_M}{\rho_N\big(v\big)}dv.
  \end{gather}
Thus, we obtain
  \begin{gather}
    \label{Phi1}
    \int^{\infty}_{v_M}{\rho_N\big(v\big)}dv=\frac{1}{2N}+\frac{\pi-\gamma}{\pi}\qty(\frac{1}{2}-\frac{M}{N}).
  \end{gather}
Then, by taking the thermodynamic limit we get
  \begin{gather}
    \label{Q}
    \int^{\infty}_{Q}{\rho\big(v\big)}dv=\qty(1-\frac{\gamma}{\pi})m.
  \end{gather}

\subsection{Fourier transformation}
We define a Fourier transformation of a function $f(x)$ as
  \begin{gather}
    \tilde{f}(\omega)=\int^{\infty}_{-\infty}f(x)e^{i\omega x}dx,
  \end{gather}
which simultaneously means
  \begin{gather}
    f(x)=\frac{1}{2\pi}\int^{\infty}_{-\infty}\tilde{f}(\omega)e^{-i\omega x}d\omega.
  \end{gather}

\subsection{The exactly solvable case: $m=0$}
\label{Appendix_rho0}

We can solve the integral equation (\ref{rho}) 
explicitly only when $m=0$.
Since it follows from Eq.~(\ref{Q}) that $Q=\infty$ for $m=0$, the integral equation becomes
  \begin{gather}
    \label{rho_0_int}
    \rho_{0}(v)=a_1(v)-\int^{\infty}_{-\infty}a_2(v-x)\rho_{0}(x)dx,
  \end{gather}
where we defined the solution as $\rho_{0}(v)$.
By using Fourier transformation on both sides, we see that Eq.~(\ref{rho_0_int}) yields
  \begin{gather}
    \tilde{\rho}_{0}(\omega)=\tilde{a}_{1}(\omega)-\tilde{a}_{2}(\omega)\tilde{\rho}_{0}(\omega)
    \\\Rightarrow
    \tilde{\rho}_{0}(\omega)=\frac{\tilde{a}_{1}(\omega)}{1+\tilde{a}_{2}(\omega)}
    =\frac{\sinh\qty(\frac{\pi}{\gamma}-1)\omega}{\sinh{\frac{\pi}{\gamma}\omega}+\sinh\qty(\frac{\pi}{\gamma}-2)\omega}
    =\frac{1}{2\cosh\omega}.
  \end{gather}
As a result, $\rho_{0}(v)$ can be calculated as
  \begin{gather}
  \label{rho3}
    \rho_{0}(v)=\frac{1}{2\pi}\int^{\infty}_{-\infty}\tilde{\rho}_{0}(\omega)e^{-i\omega v}d\omega
    =\frac{1}{2\pi}\int^{\infty}_{-\infty}\frac{e^{-i\omega v}}{2\cosh\omega}d\omega
    =\frac{1}{4\cosh\frac{\pi}{2}v}.
  \end{gather}

\subsection{Wiener-Hopf method for $m>0$ case}
\label{WH1}
In the following discussion, we consider only the lowest energy density with infinitesimal magnetization $m$, which means $Q$ is sufficiently large, but not infinite.
By dividing the integration interval in (\ref{rho}) as
  \begin{gather}
    \label{rho1-2}
    \rho(v)+\int^{\infty}_{-\infty}a_2(v-x)\rho(x)dx=a_1(v)+\int_{\abs{x}>Q}a_2(v-x)\rho(x)dx
  \end{gather}
and using Fourier transformation twice, we can extract $\rho_0(v)$ from $\rho(v)$ as follows:
  \begin{gather}
    \label{rho2}
    \rho(v)=\rho_0(v)+\int_{\abs{x}>Q}R(v-x)\rho(x)dx.
  \end{gather}
Here $\rho_{0}(v)$ is the solution at $m=0$, namely $Q=\infty$ (see section S3.\ref{Appendix_rho0}), and the integral kernel $R(v)$ (see section S3.\ref{Appendix_R}) is defined as
  \begin{gather}
    \label{R}
    R(v)=\frac{1}{2\pi}\int^{\infty}_{-\infty}e^{-i\omega v}\frac{\tilde{a}_2\qty(\omega)}{1+\tilde{a}_2\qty(\omega)}d\omega
    =\frac{1}{2\pi}\int^{\infty}_{-\infty}e^{-i\omega v}\frac{\sinh\qty(\frac{\pi}{\gamma}-2)\omega}{2\cosh\omega\sinh\qty(\frac{\pi}{\gamma}-1)\omega}d\omega
  \end{gather}
with the Fourier transform of $a_n(v)$:
  \begin{gather}
    \tilde{a}_n\qty(\omega)=\int^{\infty}_{-\infty}a_n(x)e^{i\omega v}dx=\frac{\sinh\qty(\frac{\pi}{\gamma}-n)\omega}{\sinh\frac{\pi}{\gamma}\omega}.
  \end{gather}
Now we introduce new functions
  \begin{align}
    g(v)\equiv\rho(v+Q)=g_{+}(v)+g_{-}(v),
  \end{align}
  \begin{gather}
    \label{g}
    g_{\pm}(v)\equiv\Theta\qty(\pm v)g(v),
  \end{gather}
where $\Theta\qty(v)$ is a Heaviside step function.
By substituting $v+Q$ to the argument of Eq.~(\ref{rho2}), we have
  \begin{align}
    \nonumber
    g(v)=&\rho_0\qty(v+Q)+\int^{\infty}_{Q}R\qty(v+Q-x)\rho(x)dx+\int^{-Q}_{-\infty}R\qty(v+Q-x)\rho(x)dx
    \\\nonumber
    =&\rho_0\qty(v+Q)+\int^{\infty}_{0}R\qty(v-x)\rho(x+Q)dx+\int^{0}_{-\infty}R\qty(v-x+2Q)\rho(x-Q)dx,
    \\\nonumber
    =&\rho_0\qty(v+Q)+\int^{\infty}_{0}R\qty(v-x)g(x)dx+\int^{\infty}_{0}R\qty(v+x+2Q)g(x)dx
    \nonumber
    \\\label{int}
    =&\rho_0\qty(v+Q)+\int^{\infty}_{-\infty}R\qty(v-x)g_{+}(x)dx+\int^{\infty}_{-\infty}R\qty(v+x+2Q)g_{+}(x)dx.
  \end{align}
Next we investigate behaviors of $\rho_{0}(v+Q)$ and $R(v+Q)$ in $v>0$.
From Eqs.~(\ref{rho3}) and (\ref{R}), we get
  \begin{align}
    \rho_{0}(v+Q)&=\frac{1}{2\pi}\int^{\infty}_{-\infty}\frac{e^{-i\omega (v+Q)}}{2\cosh\omega}d\omega
    =\frac{1}{2\pi}\int^{\infty}_{-\infty}\tilde{\rho}_{0}(\omega)e^{-i\omega (v+Q)}d\omega
    \\
    R(v+Q)&=\frac{1}{2\pi}\int^{\infty}_{-\infty}e^{-i\omega (v+Q)}\frac{\sinh\qty(\frac{\pi}{\gamma}-2)\omega}{2\cosh\omega\sinh\qty(\frac{\pi}{\gamma}-1)\omega}d\omega
    \\\nonumber
    &=\frac{1}{2\pi}\int^{\infty}_{-\infty}\tilde{R}(\omega)e^{-i\omega (v+Q)}d\omega.
  \end{align}
This suggests that poles of $\tilde{\rho}_{0}(\omega)$ or $\tilde{R}(\omega)$ in the lower-half plane contribute to $\rho_{0}(v+Q)$ and $R(v+Q)$, respectively.
The position of the poles can be read off from the explicit expressions for $\tilde{\rho}_0(\omega)$ and $\tilde{R}(\omega)$:
  \begin{gather}
    \label{poles1}
    \tilde{\rho}_0(\omega)=\frac{1}{2\cosh{\omega}}
    \rightarrow
    {\rm poles}:\omega=-i\pi\qty(n-\frac{1}{2}),
    \\\label{poles2}
    \tilde{R}(\omega)=\frac{\sinh{\qty(\frac{\pi}{\gamma}-2)\omega}}{2\cosh{\omega}\sinh{\qty(\frac{\pi}{\gamma}-1)\omega}}
    \rightarrow
    {\rm poles}:\omega=-i\pi\qty(n-\frac{1}{2}), -\frac{il\pi\gamma}{\pi-\gamma},
  \end{gather}
where $n, l\in\mathbb{N}$.
In the following discussion, we consider only the case where all the poles of $\tilde{R}(\omega)$ are different, in which case $\gamma\ne\pi(2n-1)/(2n-1+2l)\ (n,l\in\mathbb{N})$.
This makes all the poles of $\tilde{\rho}_0(\omega)$ and $\tilde{R}(\omega)$ simple poles, and thus we can treat them on an equal footing. 
Note that there exist double poles for $\gamma=\pi(2n-1)/(2n-1+2l)\ (n,l\in\mathbb{N})$, and we have to treat these cases separately. 
For $\gamma\ne\pi(2n-1)/(2n-1+2l)\ (n,l\in\mathbb{N})$, $\rho_{0}(v+Q)$ and $R(v+Q)$ can be written as
  \begin{align}
    \rho_{0}(v+Q)&={\rm Res}\qty(\tilde{\rho}_0,-i\frac{\pi}{2})\cdot\frac{e^{-\frac{\pi}{2}(v+Q)}}{i}
    +{\rm Res}\qty(\tilde{\rho}_0,-i\frac{3\pi}{2})\cdot\frac{e^{-\frac{3\pi}{2}(v+Q)}}{i}+\cdots,
    \\
    R(v+Q)&={\rm Res}\qty(\tilde{R},-i\frac{\pi}{2})\cdot\frac{e^{-\frac{\pi}{2}(v+Q)}}{i}
    +{\rm Res}\qty(\tilde{R},-i\frac{\pi\gamma}{\pi-\gamma})\cdot\frac{e^{-\frac{\pi\gamma}{\pi-\gamma}(v+Q)}}{i}+\cdots,
  \end{align}
for $v>0$.
Here we denoted a residue of a function $f(x)$ at $x=x_{0}$ as ${\rm Res}\qty(f,x_{0})$.
It is obvious that poles closer to the real axis contribute to the smaller power of $e^{-\qty(\pi/2)Q}$.
Therefore, Eq.~(\ref{int}) implies that $g(v)$ can also be expanded as
  \begin{gather}
    \label{exp_g}
    g(v)=g^{(1)}(v)+g^{(2)}(v)+\cdots,
  \end{gather}
where 
superscripts denote increasing powers of $e^{-\qty(\pi/2)Q}$.
By substituting Eq.~(\ref{exp_g}) into Eq.~(\ref{int}) and then comparing the terms at each order in $e^{-\qty(\pi/2)Q}$, we obtain
  \begin{gather}
    g^{(1)}(v)=\Big[\rho_0\qty(v+Q)\Big]^{(1)}+\int^{\infty}_{-\infty}R\qty(v-x)g^{(1)}_{+}(x)dx,
  \end{gather}
  \begin{align}
    g^{(2)}(v)=\Big[\rho_0\qty(v+Q)\Big]^{(2)}+\int^{\infty}_{-\infty}R\qty(v-x)g^{(2)}_{+}(x)dx
    +\Bigg[\int^{\infty}_{-\infty}R\qty(v+x+2Q)g^{(1)}_{+}(x)dx\Bigg]^{(2)},
  \end{align}
  \begin{align}
    \nonumber
    g^{(3)}(v)=\Big[\rho_0\qty(v+Q)\Big]^{(3)}+\int^{\infty}_{-\infty}R\qty(v-x)g^{(3)}_{+}(x)&dx
    \\
    +\Bigg[\int^{\infty}_{-\infty}R\qty(v+x+2Q)g^{(1)}_{+}(x)dx\Bigg]^{(3)}
    &+\Bigg[\int^{\infty}_{-\infty}R\qty(v+x+2Q)g^{(2)}_{+}(x)dx\Bigg]^{(3)},
  \end{align}
where 
superscripts again denote increasing powers of $e^{-\qty(\pi/2)Q}$. 
Each of the above equations is a linear integral equation of Wiener-Hopf type.
By using Fourier transformation, we get
 \begin{gather}
   \label{g1}
   \tilde{g}^{(1)}_{+}(\omega)+\tilde{g}^{(1)}_{-}(\omega)=\Big[\tilde{\rho}_0(\omega)e^{-i\omega Q}\Big]^{(1)}+\tilde{R}(\omega)\tilde{g}^{(1)}_{+}(\omega),
   \\
   \tilde{g}^{(2)}_{+}(\omega)+\tilde{g}^{(2)}_{-}(\omega)=\Big[\tilde{\rho}_0(\omega)e^{-i\omega Q}\Big]^{(2)}+\tilde{R}(\omega)\tilde{g}^{(2)}_{+}(\omega)+\Big[\tilde{R}(\omega)\tilde{g}^{(1)}_{+}(-\omega)e^{-i2\omega Q}\Big]^{(2)},
   \\
   \tilde{g}^{(3)}_{+}(\omega)+\tilde{g}^{(3)}_{-}(\omega)=\Big[\tilde{\rho}_0(\omega)e^{-i\omega Q}\Big]^{(3)}+\tilde{R}(\omega)\tilde{g}^{(3)}_{+}(\omega)+\Big[\tilde{R}(\omega)\tilde{g}^{(1)}_{+}(-\omega)e^{-i2\omega Q}\Big]^{(3)}+\Big[\tilde{R}(\omega)\tilde{g}^{(2)}_{+}(-\omega)e^{-i2\omega Q}\Big]^{(3)}.
 \end{gather}
 
\smallskip 
 
Now we introduce a convenient factorization (see section S3.\ref{Appendix_1-R})
 \begin{gather}
   \label{1-R=GG}
   1-\tilde{R}(\omega)=\frac{1}{G_{+}(\omega)G_{-}(\omega)},
 \end{gather}
where $G_{+}(\omega)$ and $G_{-}(\omega)$ are written as
 \begin{align}
   G_{+}(\omega)=\frac{\sqrt{2\qty(\pi-\gamma)}\Gamma\qty(1-i\frac{\omega}{\gamma})}{\Gamma\qty(\frac{1}{2}-i\frac{\omega}{\pi})\Gamma\qty(1-i\omega\frac{\pi-\gamma}{\pi\gamma})}\qty(\frac{\qty(\frac{\pi}{\gamma}-1)^{\frac{\pi}{\gamma}-1}}{\qty(\frac{\pi}{\gamma})^{\frac{\pi}{\gamma}}})^{-i\frac{\omega}{\pi}}
   =G_{-}(-\omega)
 \end{align}
and are analytic  and non-zero in the upper and lower half-plane, respectively.
They also show algebraic convergence as follows~\cite{Hamer}:
  \begin{gather}
    \label{alge1}
    G_{\pm}(\omega)\overset{|\omega|\rightarrow\infty}{\sim}1+\mathcal{O}(\omega^{-1}).
  \end{gather}
Then Eq.~(\ref{g1}) becomes
 \begin{gather}
   \label{g2}
   \frac{\tilde{g}^{(1)}_{+}(\omega)}{G_{+}(\omega)}+G_{-}(\omega)\tilde{g}^{(1)}_{-}(\omega)=G_{-}(\omega)\Big[\tilde{\rho}_0(\omega)e^{-i\omega Q}\Big]^{(1)}.
 \end{gather}
Now we recall that a Fourier transform $\tilde{f}(\omega)$ can be split as follows:
  \begin{gather}
    \tilde{f}(\omega)=\tilde{f}_{+}(\omega)+\tilde{f}_{-}(\omega),
  \end{gather}
where $\tilde{f}_{+}(\omega)$ and $\tilde{f}_{-}(\omega)$ are defined as
  \begin{gather}
    \label{def_pm}
    \tilde{f}_{\pm}(\omega)\equiv\pm \frac{i}{2\pi}\int_{-\infty}^{\infty}\frac{\tilde{f}({\omega}^{\prime})}{\omega-{\omega}^{\prime}\pm i0}d{\omega}^{\prime}
    \left(=\int^{\infty}_{-\infty}\Theta(\pm x)f(x)e^{i\omega x}dx\right)
  \end{gather}
and are analytic in the upper and lower half-plane, respectively (actually, $\tilde{g}^{(n)}_{\pm}(\omega)$ are examples).
They obviously show algebraic convergence as follows:
  \begin{gather}
    \label{alge2}
    \tilde{f}_{\pm}(\omega)\overset{|\omega|\rightarrow\infty}{\sim}\mathcal{O}(\omega^{-1}).
  \end{gather}
\\
Then Eq.~(\ref{g2}) yields
  \begin{align}
    \label{g3}
    \frac{\tilde{g}^{(1)}_{+}(\omega)}{G_{+}(\omega)}-\Big[G_{-}(\omega)\tilde{\rho}_0(\omega)e^{-i\omega Q}\Big]^{(1)}_{+}
    =&-G_{-}(\omega)\tilde{g}^{(1)}_{-}(\omega)+\Big[G_{-}(\omega)\tilde{\rho}_0(\omega)e^{-i\omega Q}\Big]^{(1)}_{-}
    \\
    \equiv&P(\omega).
  \end{align}
We see that the left- and right-hand side of Eq.~(\ref{g3}) are analytic in the upper and lower half-plane, respectively.
Since both of them are analytic on the real axis, the right-hand side of Eq.~(\ref{g3}) is the analytic continuation of the left-hand side, and thus there should be the entirely analytic form $P(\omega)$ \cite{Morse}.
However, Eqs.~(\ref{alge1}), (\ref{alge2}) and (\ref{g3}) suggest that $P(\omega)$ shows the following algebraic convergence:
  \begin{gather}
    P(\omega)\overset{|\omega|\rightarrow\infty}{\sim}\mathcal{O}(\omega^{-1}),
  \end{gather}
and therefore regularity of $P(\omega)$ leads to $P(\omega)=0$.

In the following discussion, we need only $\tilde{g}_{+}(\omega)$ for our purposes.
From the above discussion, $\tilde{g}^{(1)}_{+}(\omega)$ is written as
  \begin{gather}
    \tilde{g}^{(1)}_{+}(\omega)=G_{+}(\omega)\Big[G_{-}(\omega)\tilde{\rho}_0(\omega)e^{-i\omega Q}\Big]^{(1)}_{+},
  \end{gather}
and $\tilde{g}^{(2)}_{+}(\omega)$ and $\tilde{g}^{(3)}_{+}(\omega)$ can also be obtained in the same way
  \begin{gather}
    \tilde{g}^{(2)}_{+}(\omega)=G_{+}(\omega)\Bigg\{\Big[G_{-}(\omega)\tilde{\rho}_0(\omega)e^{-i\omega Q}\Big]^{(2)}_{+}
    +\Big[G_{-}(\omega)\tilde{R}(\omega)\tilde{g}^{(1)}_{+}(-\omega)e^{-i2\omega Q}\Big]^{(2)}_{+}\Bigg\},
    \\
    \tilde{g}^{(3)}_{+}(\omega)=G_{+}(\omega)\Bigg\{\Big[G_{-}(\omega)\tilde{\rho}_0(\omega)e^{-i\omega Q}\Big]^{(3)}_{+}
    +\Big[G_{-}(\omega)\tilde{R}(\omega)\tilde{g}^{(1)}_{+}(-\omega)e^{-i2\omega Q}\Big]^{(3)}_{+}+\Big[G_{-}(\omega)\tilde{R}(\omega)\tilde{g}^{(2)}_{+}(-\omega)e^{-i2\omega Q}\Big]^{(3)}_{+}\Bigg\},
  \end{gather}
but all the $\big[\cdots\big]_{+}$ are 
to be calculated.
The definition (\ref{def_pm}) and the existence of $e^{-in\omega Q}$ in every $\big[\cdots\big]_{+}$ mean that the power of $e^{-\qty(\pi/2)Q}$ in every term is determined by poles in the lower half-plane of $\tilde{\rho}_0(\omega)$ and $\tilde{R}(\omega)$.
By using Eqs.~(\ref{poles1}) and (\ref{poles2}), we obtain
  \begin{align}
    \nonumber
    \tilde{g}_{+}(\omega)&=\tilde{g}^{(1)}_{+}(\omega)+\tilde{g}^{(2)}_{+}(\omega)+\tilde{g}^{(3)}_{+}(\omega)+\cdots
    \\\nonumber
    &=G_{+}(\omega)\Bigg\{\frac{c_{1,1}}{\omega+i\frac{\pi}{2}}e^{-\frac{\pi Q}{2}}+\bigg(\frac{c_{2,1}}{\omega+i\frac{3\pi}{2}}
    +\frac{c_{2,2}}{\omega+i\frac{\pi}{2}}\bigg)e^{-\frac{3\pi Q}{2}}
    \\
    \label{g_exp}
    &\qquad \qquad \quad  +\frac{c_{2,3}}{\omega+i\frac{\pi\gamma}{\pi-\gamma}}e^{-\qty(\frac{\pi}{2}+\frac{2\pi\gamma}{\pi-\gamma})Q}
    +\bigg(\frac{c_{3,1}}{\omega+i\frac{5\pi}{2}}+\frac{c_{3,2}+c_{3,3}}{\omega+i\frac{\pi}{2}}\bigg)e^{-\frac{5\pi Q}{2}}
    +\cdots
    \Bigg\}
  \end{align}
where $c_{1,1}, c_{2,1}, c_{2,2}, c_{2,3},c_{3,1},,c_{3,2}$ and $c_{3,3}$ are
  \begin{align}
    \label{c1}
    c_{1,1}&=\frac{i}{2}G_{+}\qty(i\frac{\pi}{2}),
    \\
    c_{2,1}&=-\frac{i}{2}G_{+}\qty(i\frac{3\pi}{2}),
    \\
    c_{2,2}&=\frac{i}{4\pi}\tan\qty(\frac{\pi^2}{2\gamma})G_{+}^{3}\qty(i\frac{\pi}{2}),
    \\\label{c13}
    c_{2,3}&=\frac{i\gamma}{2\pi(\pi+\gamma)}\tan\qty(\frac{\pi\gamma}{\pi-\gamma})G_{+}\qty(i\frac{\pi}{2})G_{+}^{2}\qty(i\frac{\pi\gamma}{\pi-\gamma}),
    \\
    c_{3,1}&=\frac{i}{2}G_{+}\qty(i\frac{5\pi}{2}),
    \\
    c_{3,2}&=\frac{c_{2,1}}{4\pi}\tan\qty(\frac{\pi^2}{2\gamma})G_{+}^{2}\qty(i\frac{\pi}{2}),
    \\\label{c33}
    c_{3,3}&=\frac{c_{2,2}}{2\pi}\tan\qty(\frac{\pi^2}{2\gamma})G_{+}^{2}\qty(i\frac{\pi}{2}).
  \end{align}
Similarly, all the $\tilde{g}^{(n)}_{+}(\omega)$ can be evaluated by focusing on the poles.
As a result, $\tilde{g}_{+}(\omega)$ can be written as
  \begin{gather}
    \label{g_general}
    \tilde{g}_{+}(\omega)=\sum_{n=1}^{\infty}\tilde{g}^{(n)}_{+}(\omega)=e^{-\frac{\pi Q}{2}}\sum_{k,l\geq0}A_{kl}(\omega)~\qty(e^{-\frac{\pi Q}{2}})^{2k+\frac{4l\gamma}{\pi-\gamma}},
  \end{gather}
where $A_{kl}(\omega)$ are calculable coefficients depending on $\omega$ and $\gamma$.
Actually, we can calculate the exact values of these coefficients, and this is one of the beneficial points of this method.

\subsection{Derivation of $R(v)$}
\label{Appendix_R}
By using Fourier transformation on both sides of Eq.~(\ref{rho1-2}), we obtain
  \begin{gather}
    \nonumber
    \tilde{\rho}(\omega)+\tilde{a}_{2}(\omega)\tilde{\rho}(\omega)
    =\tilde{a}_{1}(\omega)+\tilde{a}_{2}(\omega)\int_{\abs{x}>Q}\rho(x)e^{i\omega x}dx
    \\
    \label{rho_omega}
    \Rightarrow
    \tilde{\rho}(\omega)=\frac{\tilde{a}_{1}(\omega)}{1+\tilde{a}_{2}(\omega)}+\frac{\tilde{a}_{2}(\omega)}{1+\tilde{a}_{2}(\omega)}\int_{\abs{x}>Q}\rho(x)e^{i\omega x}dx.
  \end{gather}
By using Fourier transformation again, we see that Eq.~(\ref{rho_omega}) yields
  \begin{align*}
    \rho(v)=&\rho_{0}(v)+\int_{\abs{x}>Q}\rho(x)dx\int^{\infty}_{-\infty}\frac{d{\omega}}{2\pi}\frac{\tilde{a}_{1}(\omega)}{1+\tilde{a}_{2}(\omega)}e^{i\omega(x-v)}
    \\
    =&\rho_0(v)+\int_{\abs{x}>Q}R(v-x)\rho(x)dx,
  \end{align*}
where we define $R(v)$ as
  \begin{gather*}
    R(v)=\frac{1}{2\pi}\int^{\infty}_{-\infty}e^{-i\omega v}\frac{\tilde{a}_2\qty(\omega)}{1+\tilde{a}_2\qty(\omega)}d\omega.
  \end{gather*}

\subsection{Decomposition of $1-\tilde{R}(\omega)$}
\label{Appendix_1-R}
Here we make some comments on the convenient factorization (\ref{1-R=GG})
  \begin{gather*}
    1-\tilde{R}(\omega)=\frac{1}{G_{+}(\omega)G_{-}(\omega)}.
  \end{gather*}
Since the Fourier transform of $R(v)$ is
  \begin{gather}
    \tilde{R}(\omega)=\frac{\tilde{a}_2\qty(\omega)}{1+\tilde{a}_2\qty(\omega)}
    =\frac{\sinh\qty(\frac{\pi}{\gamma}-2)\omega}{2\cosh\omega\sinh\qty(\frac{\pi}{\gamma}-1)\omega},
  \end{gather}
$1-\tilde{R}(\omega)$ can be written as
  \begin{gather}
    1-\tilde{R}(\omega)=\frac{1}{1+\tilde{a}_2\qty(\omega)}
    =\frac{\sinh\frac{\pi}{\gamma}\omega}{2\cosh\omega\sinh\qty(\frac{\pi}{\gamma}-1)\omega}.
  \end{gather}
Using the formulas
  \begin{gather}
    \sin{z}=\frac{z}{\Gamma\qty(1+\frac{z}{\pi})\Gamma\qty(1-\frac{z}{\pi})},
    \\
    \cos{z}=\frac{\pi}{\Gamma\qty(\frac{1}{2}+\frac{z}{\pi})\Gamma\qty(\frac{1}{2}-\frac{z}{\pi})},
  \end{gather}
we see that $1-\tilde{R}(\omega)$ factorizes into a product as
  \begin{align}
    \nonumber
    1-\tilde{R}(\omega)
    =&\frac{1}{2(\pi-\gamma)}\frac{\Gamma\qty(\frac{1}{2}-i\frac{\omega}{\pi})\Gamma\qty(\frac{1}{2}+i\frac{\omega}{\pi})\Gamma\qty(1-i\omega\frac{\pi-\gamma}{\pi\gamma})\Gamma\qty(1+i\omega\frac{\pi-\gamma}{\pi\gamma})}{\Gamma\qty(1-i\frac{\omega}{\gamma})\Gamma\qty(1+i\frac{\omega}{\gamma})}
    \\\nonumber
    =&\frac{1}{\sqrt{2(\pi-\gamma)}}\frac{\Gamma\qty(\frac{1}{2}-i\frac{\omega}{\pi})\Gamma\qty(1-i\omega\frac{\pi-\gamma}{\pi\gamma})}{\Gamma\qty(1-i\frac{\omega}{\gamma})}e^{-i\omega\psi}
    \cdot
    \frac{1}{\sqrt{2(\pi-\gamma)}}\frac{\Gamma\qty(\frac{1}{2}+i\frac{\omega}{\pi})\Gamma\qty(1+i\omega\frac{\pi-\gamma}{\pi\gamma})}{\Gamma\qty(1+i\frac{\omega}{\gamma})}e^{i\omega\psi}
    \\
    \equiv&\frac{1}{G_{+}(\omega)G_{+}(-\omega)},
  \end{align}
where we defined $G_{+}(\omega)$ as
  \begin{gather}
    G_{+}(\omega)\equiv\frac{\sqrt{2(\pi-\gamma)}\Gamma\qty(1-i\frac{\omega}{\gamma})}{\Gamma\qty(\frac{1}{2}-i\frac{\omega}{\pi})\Gamma\qty(1-i\omega\frac{\pi-\gamma}{\pi\gamma})}e^{i\omega\psi}.
  \end{gather}
Then $\psi$ is determined by imposing the condition
  \begin{gather}
    G_{\pm}(\omega)\xrightarrow{|\omega|\rightarrow\infty}1
  \end{gather}
as
  \begin{gather}
    \psi=\frac{1}{\gamma}\Bigg[{\rm ln}\qty(\frac{\pi}{\pi-\gamma})-\frac{\gamma}{\pi}{\rm ln}\qty(\frac{\gamma}{\pi-\gamma})\Bigg].
  \end{gather}
In the end, we obtain \cite{Takahashi}
  \begin{gather}
    \nonumber
    G_{+}(\omega)=\frac{\sqrt{2\qty(\pi-\gamma)}\Gamma\qty(1-i\frac{\omega}{\gamma})}{\Gamma\qty(\frac{1}{2}-i\frac{\omega}{\pi})\Gamma\qty(1-i\omega\frac{\pi-\gamma}{\pi\gamma})}\qty(\frac{\qty(\frac{\pi}{\gamma}-1)^{\frac{\pi}{\gamma}-1}}{\qty(\frac{\pi}{\gamma})^{\frac{\pi}{\gamma}}})^{-i\frac{\omega}{\pi}}
    =G_{-}(-\omega).
  \end{gather}
Since $1/\Gamma(z)$ is analytic in the entire plane, the poles of $G_{+}(\omega)$ are determined by those of $\Gamma (1-i\omega/\gamma)$, namely $\omega=-in\gamma$ ($n\in\mathbb{N}$).
Thus, $G_{\pm}(\omega)$ are analytic and non-zero in the upper and lower half-plane, respectively.

\subsection{The lowest energy density}
Here we calculate the lowest energy density of each sector $m$.
We can see that Eqs.~(\ref{e0}) and (\ref{Q}) are expressed by $g_{+}(v)$ as
  \begin{align}
    \nonumber
    e(m)-e\qty(0)
    &=2\pi A\Bigg\{\int^{\infty}_{Q}+\int^{-Q}_{-\infty}\Bigg\}\rho_{0}(x)\rho\qty(x)dx
    \\
    &=4\pi A\int^{\infty}_{-\infty}\rho_0\qty(x+Q)g_{+}\qty(x)dx
    \label{e0_g}
    =\pi A\int^{\infty}_{-\infty}\frac{g_{+}\qty(x)}{\cosh\frac{\pi}{2}\qty(x+Q)}dx,
  \end{align}
\vspace{-0.3cm}
  \begin{gather}
    \label{g0}
    \int^{\infty}_{Q}{\rho(x)}dx
    =\int^{\infty}_{-\infty}g_{+}(x)dx=\tilde{g}_{+}(0)
    =\qty(1-\frac{\gamma}{\pi})m.
  \end{gather}
Since the relations (\ref{g_exp}) and (\ref{g0}) give
  \begin{align}
    \nonumber
    \qty(1-\frac{\gamma}{\pi})m&=\tilde{g}^{(1)}_{+}(0)+\tilde{g}^{(2)}_{+}(0)+\tilde{g}^{(3)}_{+}(0)+\cdots
    \\
    &=G_{+}(0)\Bigg\{\frac{c_{1,1}}{i\frac{\pi}{2}}e^{-\frac{\pi Q}{2}}+\bigg(\frac{c_{2,1}}{i\frac{3\pi}{2}}
    +\frac{c_{2,2}}{i\frac{\pi}{2}}\bigg)e^{-\frac{3\pi Q}{2}}
    +\frac{c_{2,3}}{i\frac{\pi\gamma}{\pi-\gamma}}e^{-\qty(\frac{\pi}{2}+\frac{2\pi\gamma}{\pi-\gamma})Q}
    +\bigg(\frac{c_{3,1}}{i\frac{5\pi}{2}}+\frac{c_{3,2}}{i\frac{\pi}{2}}\bigg)e^{-\frac{5\pi Q}{2}}
    +\cdots
    \Bigg\},
  \end{align}
$e^{\pi Q/2}$ can be expanded with respect to $m$.
Then by using the relation
  \begin{align}
    \nonumber
    \int^{\infty}_{-\infty}\frac{g_{+}(x)}{\cosh\frac{\pi}{2}\qty(x+Q)}dx
    &=2\int_{-\infty}^{\infty}g_{+}(x)e^{-\frac{\pi}{2}\qty(x+Q)}\qty(1-e^{-\pi\qty(x+Q)}+\cdots)dx
    \\
    &=2e^{-\frac{\pi Q}{2}}\Bigg(\tilde{g}_{+}\bigg(i\frac{\pi}{2}\bigg)-\tilde{g}_{+}\qty(i\frac{3\pi}{2})e^{-\pi Q}+\cdots\Bigg),
  \end{align}
Eq.~(\ref{e0_g}) can be expressed by $e^{-\pi Q/2}$, which means that it can be expressed also by $m$.
As a result, we obtain
  \begin{align}
    e(m)-e\qty(0)
    &=m^2\sum_{k,l\geq 0}B_{kl}~ m^{2k+\frac{4l\gamma}{\pi-\gamma}}
    \\\nonumber
    &=\frac{\pi J\sin{\gamma}}{2\gamma(\pi-\gamma)}\cdot\frac{1}{2!}\Big(2(\pi-\gamma)m\Big)^2
    \\\nonumber
    &\quad -\frac{J\sin{\gamma}}{8\gamma(\pi-\gamma)}
    \left[
    \frac{\Gamma\big(\frac{3\pi}{2\gamma}\big){\Gamma\big(\frac{\pi-\gamma}{2\gamma}\big)}^3}{\Gamma\big(\frac{3(\pi-\gamma)}{2\gamma}\big){\Gamma\big(\frac{\pi}{2\gamma}\big)}^3}
    +
    \frac{3\pi\tan{\big(\frac{\pi^{2}}{2\gamma}\big)}}{\pi-\gamma}
    \right]\cdot\frac{1}{4!}\Big(2(\pi-\gamma)m\Big)^4
    \\\nonumber
    &\quad -\frac{\pi^{2}J\sin{\gamma}\tan\qty(\frac{\pi^{2}}{\pi-\gamma})}{8\gamma(\pi-\gamma)^{2}}
    \frac{{\Gamma\big(\frac{\pi}{\pi-\gamma}\big)}^2}{{\Gamma\big(\frac{3\pi-\gamma}{2(\pi-\gamma)}\big)}^2}
    \left[\frac{\Gamma\big(\frac{\pi-\gamma}{2\gamma}\big)}{4\sqrt{\pi}\Gamma\big(\frac{\pi}{2\gamma}\big)}\right]^{\frac{4\gamma}{\pi-\gamma}}\cdot\Big(2(\pi-\gamma)m\Big)^{\frac{2(\pi+\gamma)}{\pi-\gamma}}
    \\\nonumber
    &\quad +
    \frac{3J\sin{\gamma}}{32\pi\gamma(\pi-\gamma)}
    \left[
    \frac{\Gamma\big(\frac{5\pi}{2\gamma}\big){\Gamma\big(\frac{\pi-\gamma}{2\gamma}\big)}^5}{\Gamma\big(\frac{5(\pi-\gamma)}{2\gamma}\big){\Gamma\big(\frac{\pi}{2\gamma}\big)}^5}
    -\frac{5}{3}\cdot\frac{\Gamma\big(\frac{3\pi}{2\gamma}\big)^2{\Gamma\big(\frac{\pi-\gamma}{2\gamma}\big)}^6}{\Gamma\big(\frac{3(\pi-\gamma)}{2\gamma}\big)^2{\Gamma\big(\frac{\pi}{2\gamma}\big)}^6}
    \right.
    \\\nonumber
    &\qquad\qquad\qquad\quad  \left.+
    \frac{15\pi^2\tan^2{\big(\frac{\pi^{2}}{2\gamma}\big)}}{(\pi-\gamma)^2}
    +
    \frac{5\pi\tan{\big(\frac{\pi^{2}}{2\gamma}\big)}}{\pi-\gamma}\cdot\frac{\Gamma\big(\frac{3\pi}{2\gamma}\big){\Gamma\big(\frac{\pi-\gamma}{2\gamma}\big)}^3}{\Gamma\big(\frac{3(\pi-\gamma)}{2\gamma}\big){\Gamma\big(\frac{\pi}{2\gamma}\big)}^3}
    \right]\cdot\frac{1}{6!}\Big(2(\pi-\gamma)m\Big)^6
    \\\label{e0(m)}
    &\quad +\big(\rm higher\ order\ terms\big),
  \end{align}
 where $B_{kl}$ are calculable coefficients depending on $\gamma$.
  
\subsection{Calculation of nonlinear Drude weights}
In order to obtain the Drude weights, we have to introduce the $U(1)$ flux to the above discussions.
The new Hamiltonian without the magnetic field is defined as
  \begin{gather}
    \hat{\mathcal{H}}(\Theta)
    =\sum_{l=1}^{N}2J\bigg[\frac{1}{2}e^{i\frac{\Theta}{N}}\hat{S}_{l}^{+}\hat{S}_{l+1}^{-}+{\rm h.c.}+\Delta \hat{S}_{l}^{z}\hat{S}_{l+1}^{z}\bigg].
  \end{gather}
As we have already discussed, the Hamiltonian of this case can be regarded as the original chain with the DM interaction (\ref{deriv_DM}).
Then the Bethe equations are modified as
    \begin{gather}
      \label{Z_Theta}
      \mathcal{Z}_N\big(v_j\qty(\Theta),\Theta\big)=\frac{2\pi I_j}{N}=\frac{\pi}{N}\left(-M+2j-1\right)
      \quad \left(j=1,2,\cdots,M \right),
    \end{gather}
  where
    \begin{gather}
      \mathcal{Z}_N\big(v,\Theta\big)\equiv p_1(v)+\frac{\Theta}{N}-\frac{1}{N}\sum_{k=1}^{M}{p_2\big(v-v_k\qty(\Theta)\big)}.
    \end{gather}
Since we have
  \begin{align}
    \nonumber
    \mathcal{Z}_N\big(\infty,\Theta\big)-\mathcal{Z}_N\big(v_M\qty(\Theta),\Theta\big)
    =&\qty(\pi-\gamma+\frac{\Theta}{N}-\big(\pi-2\gamma\big)\frac{M}{N})-\frac{\pi}{N}\big(M-1\big)
    \\
    =&\frac{\pi}{N}+2\qty(\pi-\gamma)\qty(\frac{1}{2}-\frac{M}{N})+\frac{\Theta}{N},
  \end{align}
  \begin{align}
    \nonumber
    \mathcal{Z}_N\big(v_1\qty(\Theta),\Theta\big)-\mathcal{Z}_N\big(-\infty,\Theta\big)
    =&-\frac{\pi}{N}\big(M-1\big)-\qty(-(\pi-\gamma)+\frac{\Theta}{N}+\big(\pi-2\gamma\big)\frac{M}{N})
    \\
    =&\frac{\pi}{N}+2\qty(\pi-\gamma)\qty(\frac{1}{2}-\frac{M}{N})-\frac{\Theta}{N},
  \end{align}
the Bethe roots are uniquely determined and the set of real solutions $\{v_j(\Theta)\}$ satisfy $-\infty\leq v_1(\Theta)<v_2(\Theta)<\ldots<v_M(\Theta)\leq\infty$ under the condition that
  \begin{gather}
    \mathcal{Z}_N\big(\infty,\Theta\big)-\mathcal{Z}_N\big(v_M\qty(\Theta),\Theta\big)\geq0\land
    \mathcal{Z}_N\big(v_1\qty(\Theta),\Theta\big)-\mathcal{Z}_N\big(-\infty,\Theta\big)\geq 0,
  \end{gather}
which reduces to
  \begin{gather}
    |\Theta|\leq\pi+2N\qty(\pi-\gamma)\qty(\frac{1}{2}-\frac{M}{N}).
  \end{gather}
By changing the sign of $\Theta$ in the Bethe equations (\ref{Z_Theta}), we get
  \begin{gather}
    \nonumber
    p_1\big(v_j\qty(-\Theta)\big)-\frac{\Theta}{N}-\frac{1}{N}\sum_{k=1}^{M}{p_2\big(v_j\qty(-\Theta)-v_k\qty(-\Theta)\big)}=\frac{2\pi I_j}{N}
    \\
    \Leftrightarrow
    p_1\big(-v_j\qty(-\Theta)\big)+\frac{\Theta}{N}-\frac{1}{N}\sum_{k=1}^{M}{p_2\big(-v_j\qty(-\Theta)+v_k\qty(-\Theta)\big)}
    =-\frac{2\pi I_j}{N}
    =\frac{2\pi I_{M-j+1}}{N},
  \end{gather}
and thus the uniqueness of $\{v_j(\Theta)\}$ leads to
  \begin{gather}
    \label{root}
    -v_j\qty(-\Theta)=v_{M-j+1}\qty(\Theta).
  \end{gather}
Then we define the energy density $e(M,\Theta)$ calculated from these roots as
  \begin{align}
    e(M,\Theta)=&-\frac{2\pi A}{N}\sum_{j=1}^{M}a_1\big(v_j(\Theta)\big)+\frac{\Delta}{2}
    \\
    =&\frac{1}{N}\sum_{j=1}^{M}\frac{2J\sin^2{\gamma}}{\cos{\gamma}-\cosh{\gamma {v_{j}(\Theta)}}}+\frac{\Delta}{2},
  \end{align}
and thus the relation (\ref{root}) gives
  \begin{gather}
  \label{e0_sym}
    e(M,\Theta)=e(M,-\Theta).
  \end{gather}

\smallskip

In the case $|\Theta|\leq\pi$, $e(M,\Theta)$ corresponds to the lowest energy density in the sector of $M$. Otherwise, $e(M,\Theta)$ corresponds to the excited energy density in the same sector.
Now by introducing the function $\rho_N(v,\Theta)$ as
  \begin{gather}
    \rho_N\big(v,\Theta\big)\equiv\frac{1}{2\pi}\dv{\mathcal{Z}_N\big(v,\Theta\big)}{v},
  \end{gather}
we obtain
  \begin{gather}
    \label{rho_N_theta}
    \rho_N\big(v,\Theta\big)=a_1(v)-\frac{1}{N}\sum_{k=1}^{M}{a_2\big(v-v_k\qty(\Theta)\big)},
    \\
    \int^{\infty}_{v_M\qty(\Theta)}{\rho_N\big(v,\Theta\big)}dv=\frac{1}{2N}+\frac{\pi-\gamma}{\pi}\qty(\frac{1}{2}-\frac{M}{N})+\frac{\Theta}{2\pi N},
    \\
    \int^{v_1\qty(\Theta)}_{-\infty}{\rho_N\big(v,\Theta\big)}dv=\frac{1}{2N}+\frac{\pi-\gamma}{\pi}\qty(\frac{1}{2}-\frac{M}{N})-\frac{\Theta}{2\pi N}.
  \end{gather}
We now introduce $\theta\equiv\Theta/N$ and $m\equiv1/2-M/N$. In the thermodynamic limit, we get the following relations for $|\theta|<2(\pi-\gamma)m$:
  \begin{gather}
    \label{e0_m_theta}
    e\big(m,\theta\big)=-2\pi A\int^{Q^{(+)}\qty(\theta)}_{-Q^{(-)}\qty(\theta)}a_1(x)\rho\big(x,\theta\big)dx+\frac{\Delta}{2}=e\big(m,-\theta\big),
    \\\label{rho_theta}
    \rho\big(v,\theta\big)=a_1(v)-\int^{Q^{(+)}\qty(\theta)}_{-Q^{(-)}\qty(\theta)}a_2(v-x)\rho\big(x,\theta\big)dx,
    \\\label{rho_+theta}
    \int^{\infty}_{Q^{(+)}\qty(\theta)}{\rho\big(v,\theta\big)}dv=\qty(1-\frac{\gamma}{\pi})m+\frac{\theta}{2\pi},
    \\\label{rho_-theta}
    \int^{-Q^{(-)}\qty(\theta)}_{-\infty}{\rho\big(v,\theta\big)}dv=\qty(1-\frac{\gamma}{\pi})m-\frac{\theta}{2\pi},
  \end{gather}
where $e(m,\theta),-Q^{(-)}\qty(\theta),Q^{(+)}\qty(\theta)$, and $\rho(v,\theta)$ are new representations of $e(M,\Theta),v_1\qty(\Theta),v_M\qty(\Theta)$ and $\rho_N(v,\Theta)$ in the limit, respectively.
Note that Eq.~(\ref{root}) implies  $Q^{(+)}\qty(\theta)=Q^{(-)}\qty(-\theta)$.

\smallskip

Now we consider the infinitesimal $m$ and $\theta$.
By using the Wiener-Hopf method (see the next section), we obtain the following expansion of $e(m,\theta)$ for $|\theta|<2(\pi-\gamma)m$:
  \begin{align}
    \nonumber
    e\big(m,\theta\big)-e\qty(0,0)
    =\sum_{1\leq k+l< 2\gamma/(\pi-\gamma)+1}C_{kl}~ &{\bigg[\qty(1-\frac{\gamma}{\pi})m+\frac{\theta}{2\pi}\bigg]}^{2k}{\bigg[\qty(1-\frac{\gamma}{\pi})m-\frac{\theta}{2\pi}\bigg]}^{2l}
    \\\label{exp_theta}
    &+{\mathcal{O}\scriptstyle
    \qty(\Big(\big[\qty(1-\frac{\gamma}{\pi})m+\frac{\theta}{2\pi}\big]\big[\qty(1-\frac{\gamma}{\pi})m-\frac{\theta}{2\pi}\big]\Big)^{\frac{2\gamma}{\pi-\gamma}+1})
    },
  \end{align}
where we have assumed that $4\gamma/(\pi-\gamma)$ is {\it noninteger}, and $C_{kl}$ are calculable coefficients depending on $\gamma$ and satisfying $C_{kl}=C_{lk}$ because of Eq.~(\ref{e0_sym}).
It is obvious that substitution of $\theta=0$ into the above restores Eq.~(\ref{e0(m)}).
Therefore, all the Drude weights can be calculated as
  \begin{gather}
    \label{Dcalc}
    \mathcal{D}^{(n)}=\lim_{m\rightarrow 0}\lim_{\theta\rightarrow 0}\pdv[n+1]{\theta}e(m,\theta),
  \end{gather}
and this results in
  \begin{gather}
    \mathcal{D}^{(n)}=\pdv[n+1]{e(m,0)}{\big(2(\pi-\gamma)m\big)}\Bigr|_{m=0}=\pdv[n+1]{e(m)}{\big(2(\pi-\gamma)m\big)}\Bigr|_{m=0}
  \end{gather}
when $e(m)$ is differentiable at the origin.
Note that the order of the two limits in Eq.~(\ref{Dcalc}) cannot be exchanged because of the condition $|\theta|<2(\pi-\gamma)m$.
Thus we can calculate the nonlinear Drude weights from the series expansion of the lowest energy density of each sector with respect to $m$, i.e., Eq.~(\ref{e0(m)}).
As a result, we get
  \begin{align}
     \mathcal{D}^{(1)}=\pdv[2]{e(m)}{\big(2(\pi-\gamma)m\big)}\Bigr|_{m=0}
     &=\frac{\pi J \sin{\gamma}}{2\gamma(\pi-\gamma)},
     \\
     \mathcal{D}^{(3)}=\pdv[4]{e(m)}{\big(2(\pi-\gamma)m\big)}\Bigr|_{m=0}
     &=-\frac{J\sin{\gamma}}{8\gamma(\pi-\gamma)}
    \left[
    \frac{\Gamma\big(\frac{3\pi}{2\gamma}\big){\Gamma\big(\frac{\pi-\gamma}{2\gamma}\big)}^3}{\Gamma\big(\frac{3(\pi-\gamma)}{2\gamma}\big){\Gamma\big(\frac{\pi}{2\gamma}\big)}^3}
    +
    \frac{3\pi\tan{\big(\frac{\pi^{2}}{2\gamma}\big)}}{\pi-\gamma}
    \right],
     \\\nonumber
     \mathcal{D}^{(5)}=\pdv[6]{e(m)}{\big(2(\pi-\gamma)m\big)}\Bigr|_{m=0}
     &=\frac{3J\sin{\gamma}}{32\pi\gamma(\pi-\gamma)}
    \left[
    \frac{\Gamma\big(\frac{5\pi}{2\gamma}\big){\Gamma\big(\frac{\pi-\gamma}{2\gamma}\big)}^5}{\Gamma\big(\frac{5(\pi-\gamma)}{2\gamma}\big){\Gamma\big(\frac{\pi}{2\gamma}\big)}^5}
    -
    \frac{5}{3}\cdot\frac{\Gamma\big(\frac{3\pi}{2\gamma}\big)^2{\Gamma\big(\frac{\pi-\gamma}{2\gamma}\big)}^6}{\Gamma\big(\frac{3(\pi-\gamma)}{2\gamma}\big)^2{\Gamma\big(\frac{\pi}{2\gamma}\big)}^6}
    \right.
    \\
    &
    \qquad\qquad\qquad\quad
    \left.
    +
    \frac{15\pi^2\tan^2{\big(\frac{\pi^{2}}{2\gamma}\big)}}{(\pi-\gamma)^2}
    +
    \frac{5\pi\tan{\big(\frac{\pi^{2}}{2\gamma}\big)}}{\pi-\gamma}\cdot\frac{\Gamma\big(\frac{3\pi}{2\gamma}\big){\Gamma\big(\frac{\pi-\gamma}{2\gamma}\big)}^3}{\Gamma\big(\frac{3(\pi-\gamma)}{2\gamma}\big){\Gamma\big(\frac{\pi}{2\gamma}\big)}^3}
    \right],
  \end{align}
in the limited regions determined by $2(\pi+\gamma)/(\pi-\gamma)>n+1$ for $\mathcal{D}^{(n)}$, where differential coefficients are well-defined at the origin $m=0$.
The above results for $\mathcal{D}^{(1)}$ and $\mathcal{D}^{(3)}$ are consistent with the previous results \cite{Sutherland,Watanabe-Oshikawa}. 

\subsection{Derivation of (\ref{exp_theta})}
The derivation of Eq.~(\ref{exp_theta}) is similar to that of Eq.~(\ref{e0(m)}). However, it is more complicated because of the presence of $U(1)$ flux. 
Let us define the functions 
  \begin{gather}
    g\qty(v,\theta)\equiv\rho\qty(v+Q^{(+)}\big(\theta\big),\theta)=g_{+}\qty(v,\theta)+g_{-}\qty(v,\theta),
    \\
    g_{\pm}\qty(v,\theta)\equiv\Theta\qty(\pm v)g\qty(v,\theta),
  \end{gather}
where $\Theta\qty(v)$ is a Heaviside step function.
Then from Eqs.~(\ref{e0_m_theta}), (\ref{rho_theta}) and (\ref{rho_+theta}) we obtain
  \begin{align}
    \nonumber
    e\big(m,\theta\big)-e\qty(0,0)
    =&2\pi A\Bigg\{\int^{\infty}_{Q^{(+)}\qty(\theta)}+\int^{-Q^{(-)}\qty(\theta)}_{-\infty}\Bigg\}\rho_{0}(x)\rho\qty(x,\theta)dx
    \\\nonumber
    =&2\pi A\Bigg\{\int^{\infty}_{-\infty}\rho_0\qty(x+Q^{(+)}\big(\theta\big))g_{+}\qty(x,\theta)dx
    +\int^{\infty}_{-\infty}\rho_0\qty(x+Q^{(-)}\big(\theta\big))g_{+}\qty(x,-\theta)dx\Bigg\}
    \\\label{e0_m_theta2}
    =&\frac{\pi A}{2}\Bigg\{\int^{\infty}_{-\infty}\frac{g_{+}\qty(x,\theta)}{\cosh\frac{\pi}{2}\qty(x+Q^{(+)}(\theta))}dx
    +\int^{\infty}_{-\infty}\frac{g_{+}\qty(x,-\theta)}{\cosh\frac{\pi}{2}\qty(x+Q^{(-)}(\theta))}dx\Bigg\},
  \end{align}
  \begin{gather}
    \label{rho_v_theta}
    \rho\qty(v,\theta)=\rho_{0}(v)+\Bigg\{\int^{\infty}_{Q^{(+)}\qty(\theta)}+\int^{-Q^{(-)}\qty(\theta)}_{-\infty}\Bigg\}R(v-x)\rho\qty(x,\theta)dx,
  \end{gather}
  \begin{gather}
    \label{g0_2}
    \int^{\infty}_{Q^{(+)}\qty(\theta)}{\rho\big(x,\theta\big)}dx
    =\int^{\infty}_{-\infty}g_{+}(x,\theta)dx=\tilde{g}_{+}(0,\theta)
    =\qty(1-\frac{\gamma}{\pi})m+\frac{\theta}{2\pi}.
  \end{gather}
Here we used the relation
  \begin{gather}
    \rho(x,-\theta)=\rho(-x,\theta),
  \end{gather}
which can be derived from Eq.~(\ref{rho_N_theta}) as follows:
  \begin{align}
    \rho_N\big(v,-\Theta\big)&=a_1(v)-\frac{1}{N}\sum_{k=1}^{M}{a_2\big(v-v_k\qty(-\Theta)\big)}
    =a_1(v)-\frac{1}{N}\sum_{k=1}^{M}{a_2\big(v+v_k\qty(\Theta)\big)}
    \\
    &=a_1(-v)-\frac{1}{N}\sum_{k=1}^{M}{a_2\big(-v-v_k\qty(\Theta)\big)}=\rho_N\big(-v,\Theta\big).
  \end{align}
By substituting $v+Q^{(+)}\qty(\theta)$ to the argument of Eq.~(\ref{rho_v_theta}), we find
  \begin{gather}
    \label{g_theta}
    g\qty(v,\theta)=\rho_0\qty(v+Q^{(+)}\big(\theta\big))+\int^{\infty}_{-\infty}R\qty(v-x)g_{+}(x,\theta)dx
    +\int^{\infty}_{-\infty}R\qty(v+x+Q^{(+)}\big(\theta\big)+Q^{(-)}\big(\theta\big))g_{+}(x,-\theta)dx.
  \end{gather}
Again we expand $g\qty(v,\theta)$ as
\begin{gather}
  \label{g_theta_exp}
  g\qty(v,\theta)=g^{(1)}\qty(v,\theta)+g^{(2)}\qty(v,\theta)+\cdots,
\end{gather}
where 
superscripts denote increasing powers of $e^{-\qty(\pi/2)Q^{(\pm)}\qty(\theta)}$.
Then substitution of Eq.~(\ref{g_theta_exp}) into Eq.~(\ref{g_theta}) gives
  \begin{gather}
      g^{(1)}\qty(v,\theta)=\bigg[\rho_0\qty(v+Q^{(+)}\big(\theta\big))\bigg]^{(1)}+\int^{\infty}_{-\infty}R\qty(v-x)g^{(1)}_{+}\qty(x,\theta)dx,
  \end{gather}
  \begin{align}
    g^{(2)}\qty(v,\theta)=\bigg[\rho_0\qty(v+Q^{(+)}\big(\theta\big))\bigg]^{(2)}+&\int^{\infty}_{-\infty}R\qty(v-x)g^{(2)}_{+}\qty(x,\theta)dx
    +\Bigg[\int^{\infty}_{-\infty}R\qty(v+x+Q^{(+)}\big(\theta\big)+Q^{(-)}\big(\theta\big))g^{(1)}_{+}\qty(x,-\theta)dx\Bigg]^{(2)},
  \end{align}
  \begin{align}
    \nonumber
    g^{(3)}\qty(v,\theta)=\bigg[\rho_0\qty(v+Q^{(+)}\big(\theta\big))\bigg]^{(3)}+\int^{\infty}_{-\infty}R\qty(v-x)g^{(3)}_{+}\qty(x,\theta)dx
    +&\Bigg[\int^{\infty}_{-\infty}R\qty(v+x+Q^{(+)}\big(\theta\big)+Q^{(-)}\big(\theta\big))g^{(1)}_{+}\qty(x,-\theta)dx\Bigg]^{(3)}
    \\
    +&\Bigg[\int^{\infty}_{-\infty}R\qty(v+x+Q^{(+)}\big(\theta\big)+Q^{(-)}\big(\theta\big))g^{(2)}_{+}\qty(x,-\theta)dx\Bigg]^{(3)}.
  \end{align}
By using Fourier transformation, we get
  \begin{gather}
    \tilde{g}^{(1)}_{+}\qty(\omega,\theta)=G_{+}(\omega)\Big[G_{-}(\omega)\tilde{\rho}_0(\omega)e^{-i\omega Q^{(+)}\qty(\theta)}\Big]^{(1)}_{+},
  \end{gather}
  \begin{align}
    \quad \tilde{g}^{(2)}_{+}\qty(\omega,\theta)=G_{+}(\omega)\Bigg\{&\Big[G_{-}(\omega)\tilde{\rho}_0(\omega)e^{-i\omega Q^{(+)}\qty(\theta)}\Big]^{(2)}_{+}
    +\bigg[G_{-}(\omega)\tilde{R}(\omega)\tilde{g}^{(1)}_{+}\qty(-\omega,-\theta)e^{-i\omega \big(Q^{(+)}(\theta)+Q^{(-)}(\theta)\big)}\bigg]^{(2)}_{+}\Bigg\},
  \end{align}
  \begin{align}
    \nonumber
    \quad \tilde{g}^{(3)}_{+}\qty(\omega,\theta)
    =G_{+}(\omega)\Bigg\{\Big[G_{-}(\omega)\tilde{\rho}_0(\omega)e^{-i\omega Q^{(+)}\qty(\theta)}\Big]^{(3)}_{+}
    &+\bigg[G_{-}(\omega)\tilde{R}(\omega)\tilde{g}^{(1)}_{+}\qty(-\omega,-\theta)e^{-i\omega \big(Q^{(+)}(\theta)+Q^{(-)}(\theta)\big)}\bigg]^{(3)}_{+}
    \\
    &+\bigg[G_{-}(\omega)\tilde{R}(\omega)\tilde{g}^{(2)}_{+}\qty(-\omega,-\theta)e^{-i\omega \big(Q^{(+)}(\theta)+Q^{(-)}(\theta)\big)}\bigg]^{(3)}_{+}\Bigg\}.
  \end{align}
Then an explicit calculation of all the $\big[\cdots\big]_{+}$ leads to
  \begin{align}
    \nonumber
    \tilde{g}_{+}(\omega,\theta)&=\tilde{g}^{(1)}_{+}(\omega,\theta)+\tilde{g}^{(2)}_{+}(\omega,\theta)+\tilde{g}^{(3)}_{+}(\omega,\theta)+\cdots
    \\\nonumber
    &=G_{+}(\omega)\Bigg\{\frac{c_{1,1}}{\omega+i\frac{\pi}{2}}e^{-\frac{\pi}{2}Q^{(+)}\qty(\theta)}+\frac{c_{2,1}}{\omega+i\frac{3\pi}{2}}e^{-\frac{3\pi}{2}Q^{(+)}\qty(\theta)}
    +\frac{c_{2,2}}{\omega+i\frac{\pi}{2}}e^{-\frac{\pi}{2}\qty(Q^{(+)}(\theta)+2Q^{(-)}\qty(\theta))}
    \\\nonumber
    &\qquad \qquad \qquad \qquad   +\frac{c_{2,3}}{\omega+i\frac{\pi\gamma}{\pi-\gamma}}e^{-\frac{\pi}{2}Q^{(-)}\qty(\theta)-\frac{\pi\gamma}{\pi-\gamma}\qty(Q^{(+)}\qty(\theta)+Q^{(-)}\qty(\theta))}
    +\frac{c_{3,1}}{\omega+i\frac{5\pi}{2}}e^{-\frac{5\pi}{2}Q^{(+)}\qty(\theta)}
    \\\label{g_exp2}
    &\qquad \qquad \qquad \qquad  +\frac{c_{3,2}}{\omega+i\frac{\pi}{2}}e^{-\frac{\pi }{2}\qty(Q^{(+)}(\theta)+4Q^{(-)}\qty(\theta))}+\frac{c_{3,3}}{\omega+i\frac{\pi}{2}}e^{-\frac{\pi}{2}\qty(3Q^{(+)}(\theta)+2Q^{(-)}\qty(\theta))}
    +\cdots
    \Bigg\},
  \end{align}
where coefficients $c_{1,1}, c_{2,1}, c_{2,2}, c_{2,3},c_{3,1},,c_{3,2}$ and $c_{3,3}$ are same as before: (\ref{c1}-\ref{c33}).
Similarly all the $\tilde{g}^{(n)}_{+}(\omega,\theta)$ can be evaluated, and as a result we obtain
  \begin{align}
    \tilde{g}_{+}(\omega,\theta)
    =\sum_{n=1}^{\infty}\tilde{g}^{(n)}_{+}(\omega,\theta)
    =X\sum_{k,l,s\geq0}D_{kls}(\omega)~ X^{2k}Y^{2l}\big(XY\big)^{\frac{2s\gamma}{\pi-\gamma}}
    +Y\sum_{k,l\geq0,\,r\geq1}E_{klr}(\omega)~ X^{2k}Y^{2l}\big(XY\big)^{\frac{2r\gamma}{\pi-\gamma}}
  \end{align}
where $D_{kls}(\omega)$ and $E_{klr}(\omega)$ are calculable coefficients depending on $\omega$ and $\gamma$, and we denoted $e^{-\qty(\pi/2)Q^{(+)}\qty(\theta)}$ and $e^{-\qty(\pi/2)Q^{(-)}\qty(\theta)}$ as $X$ and $Y$, respectively.
Therefore the relations (\ref{g0_2}) and (\ref{g_exp2}) give
  \begin{align}
    \nonumber
    \qty(1-\frac{\gamma}{\pi})m+\frac{\theta}{2\pi}
    &=\tilde{g}_{+}(0,\theta)
    \\\label{expXY}
    &=X\sum_{k,l,s\geq0}D_{kls}(0)~ X^{2k}Y^{2l}\big(XY\big)^{\frac{2s\gamma}{\pi-\gamma}}
    +Y\sum_{k,l\geq0,\,r\geq1}E_{klr}(0)~ X^{2k}Y^{2l}\big(XY\big)^{\frac{2r\gamma}{\pi-\gamma}}
    \\\nonumber
    &=G_{+}(0)\Bigg\{\frac{c_{1,1}}{i\frac{\pi}{2}}e^{-\frac{\pi}{2}Q^{(+)}\qty(\theta)}+\frac{c_{2,1}}{i\frac{3\pi}{2}}e^{-\frac{3\pi}{2}Q^{(+)}\qty(\theta)}
    +\frac{c_{2,2}}{i\frac{\pi}{2}}e^{-\frac{\pi}{2}\qty(Q^{(+)}(\theta)+2Q^{(-)}\qty(\theta))}
    \\\nonumber
    &\qquad \qquad \qquad \qquad   +\frac{c_{2,3}}{i\frac{\pi\gamma}{\pi-\gamma}}e^{-\frac{\pi}{2}Q^{(-)}\qty(\theta)-\frac{\pi\gamma}{\pi-\gamma}\qty(Q^{(+)}\qty(\theta)+Q^{(-)}\qty(\theta))}
    +\frac{c_{3,1}}{i\frac{5\pi}{2}}e^{-\frac{5\pi}{2}Q^{(+)}\qty(\theta)}
    \\
    &\qquad \qquad \qquad \qquad  +\frac{c_{3,2}}{i\frac{\pi}{2}}e^{-\frac{\pi }{2}\qty(Q^{(+)}(\theta)+4Q^{(-)}\qty(\theta))}+\frac{c_{3,3}}{i\frac{\pi}{2}}e^{-\frac{\pi}{2}\qty(3Q^{(+)}(\theta)+2Q^{(-)}\qty(\theta))}
    +\cdots
    \Bigg\}.
  \end{align}
Similarly, we have
  \begin{align}
    \nonumber
    \qty(1-\frac{\gamma}{\pi})m-\frac{\theta}{2\pi}
    &=\tilde{g}_{+}(0,-\theta)
    \\\label{expYX}
    &=Y\sum_{k,l,s\geq0}D_{kls}(0)~ Y^{2k}X^{2l}\big(XY\big)^{\frac{2s\gamma}{\pi-\gamma}}
    +X\sum_{k,l\geq0,\,r\geq1}E_{klr}(0)~ Y^{2k}X^{2l}\big(XY\big)^{\frac{2r\gamma}{\pi-\gamma}}
    \\\nonumber
    &=G_{+}(0)\Bigg\{\frac{c_{1,1}}{i\frac{\pi}{2}}e^{-\frac{\pi}{2}Q^{(-)}\qty(\theta)}+\frac{c_{2,1}}{i\frac{3\pi}{2}}e^{-\frac{3\pi}{2}Q^{(-)}\qty(\theta)}
    +\frac{c_{2,2}}{i\frac{\pi}{2}}e^{-\frac{\pi}{2}\qty(Q^{(-)}(\theta)+2Q^{(+)}\qty(\theta))}
    \\\nonumber
    &\qquad \qquad \qquad \qquad   +\frac{c_{2,3}}{i\frac{\pi\gamma}{\pi-\gamma}}e^{-\frac{\pi}{2}Q^{(+)}\qty(\theta)-\frac{\pi\gamma}{\pi-\gamma}\qty(Q^{(+)}\qty(\theta)+Q^{(-)}\qty(\theta))}
    +\frac{c_{3,1}}{i\frac{5\pi}{2}}e^{-\frac{5\pi}{2}Q^{(-)}\qty(\theta)}
    \\
    &\qquad \qquad \qquad \qquad  +\frac{c_{3,2}}{i\frac{\pi}{2}}e^{-\frac{\pi }{2}\qty(Q^{(-)}(\theta)+4Q^{(+)}\qty(\theta))}+\frac{c_{3,3}}{i\frac{\pi}{2}}e^{-\frac{\pi}{2}\qty(3Q^{(-)}(\theta)+2Q^{(+)}\qty(\theta))}
    +\cdots
    \Bigg\}.
  \end{align}
Since the relations (\ref{expXY}) and (\ref{expYX}) mean
  \begin{align}
    X=&\frac{1}{D_{kls}(0)}\left[\qty(1-\frac{\gamma}{\pi})m+\frac{\theta}{2\pi}-X\sum_{\substack{k,l,s\geq0\\k+l+s\geq1}}D_{kls}(0)~ X^{2k}Y^{2l}\big(XY\big)^{\frac{2s\gamma}{\pi-\gamma}}
    -Y\sum_{k,l\geq0,\,r\geq1}E_{klr}(0)~ X^{2k}Y^{2l}\big(XY\big)^{\frac{2r\gamma}{\pi-\gamma}}\right],
    \\
    Y=&\frac{1}{D_{kls}(0)}\left[\qty(1-\frac{\gamma}{\pi})m-\frac{\theta}{2\pi}-Y\sum_{\substack{k,l,s\geq0\\k+l+s\geq1}}D_{kls}(0)~ Y^{2k}X^{2l}\big(XY\big)^{\frac{2s\gamma}{\pi-\gamma}}
    -X\sum_{k,l\geq0,\,r\geq1}E_{klr}(0)~ Y^{2k}X^{2l}\big(XY\big)^{\frac{2r\gamma}{\pi-\gamma}}\right],
  \end{align}
sequential substitution of their right sides into $X$ and $Y$ makes it clear that $X$ and $Y$, namely $e^{-\qty(\pi/2)Q^{(\pm)}\qty(\theta)}$, can be expanded with respect to products of $\qty(1-\frac{\gamma}{\pi})m+\frac{\theta}{2\pi}$ and $\qty(1-\frac{\gamma}{\pi})m-\frac{\theta}{2\pi}$.
Then by using the relation
  \begin{align}
    \nonumber
    \int^{\infty}_{-\infty}\frac{g_{+}(x,\theta)}{\cosh\frac{\pi}{2}\qty(x+Q^{(+)}\qty(\theta))}dx
    &=2\int_{-\infty}^{\infty}g_{+}(x,\theta)e^{-\frac{\pi}{2}\qty(x+Q^{(+)}(\theta))}\qty(1-e^{-\pi\qty(x+Q^{(+)}(\theta))}+\cdots)dx
    \\
    &=2e^{-\frac{\pi}{2}Q^{(+)}(\theta)}\Bigg(\tilde{g}_{+}\bigg(i\frac{\pi}{2},\theta\bigg)-\tilde{g}_{+}\qty(i\frac{3\pi}{2},\theta)e^{-\pi Q^{(+)}(\theta)}+\cdots\Bigg),
  \end{align}
Eq.~(\ref{e0_m_theta2}) can be expressed by $e^{-(\pi/2)Q^{(\pm)}(\theta)}$, which means that it can also be expressed by products of $\qty(1-\frac{\gamma}{\pi})m+\frac{\theta}{2\pi}$ and $\qty(1-\frac{\gamma}{\pi})m-\frac{\theta}{2\pi}$.
As a result, we obtain
  \begin{align}
    \nonumber
    e\big(m,\theta\big)-e\qty(0,0)
    =\sum_{1\leq k+l< 2\gamma/(\pi-\gamma)+1}C_{kl}~ &{\bigg[\qty(1-\frac{\gamma}{\pi})m+\frac{\theta}{2\pi}\bigg]}^{2k}{\bigg[\qty(1-\frac{\gamma}{\pi})m-\frac{\theta}{2\pi}\bigg]}^{2l}
    \\
    &+{\mathcal{O}\scriptstyle
    \qty(\Big(\big[\qty(1-\frac{\gamma}{\pi})m+\frac{\theta}{2\pi}\big]\big[\qty(1-\frac{\gamma}{\pi})m-\frac{\theta}{2\pi}\big]\Big)^{\frac{2\gamma}{\pi-\gamma}+1})
    },
  \end{align}
where we have assumed that $4\gamma/(\pi-\gamma)$ is {\it noninteger}.
Although all the $C_{kl}$ are, in principle, calculable, we do not need their explicit values for our purposes.


\section*{S4. Drude weights under magnetic fields}

In the main text, we show the third-order Drude weight under the magnetic field. Here, we show the linear Drude weight and the other nonlinear Drude weights under the magnetic field. The numerical results for $\mathcal{D}_{N=800}^{(n)}(0,h)$ ($n=1,5$) and $\mathcal{D}_{N=800}^{(n)}(\Theta=0.1,h)$ ($n=2,4$) are shown in Figs.~S1 and S2, respectively. Some of the values around $\Delta=-1$ reach zero. It is natural because the gapped regime comes into $|\Delta|<1$ under the magnetic field~\cite{Takahashi}.
In terms of the NLDWs, the values are suppressed for the $\Delta$ around $1$. It seems that the divergent behavior is suppressed by the magnetic field. This behavior seems to be the same as the third-order Drude weights. The origin of this suppression is discussed in the main text.

\begin{figure*}[htbp]
  \includegraphics[width=0.865\hsize]{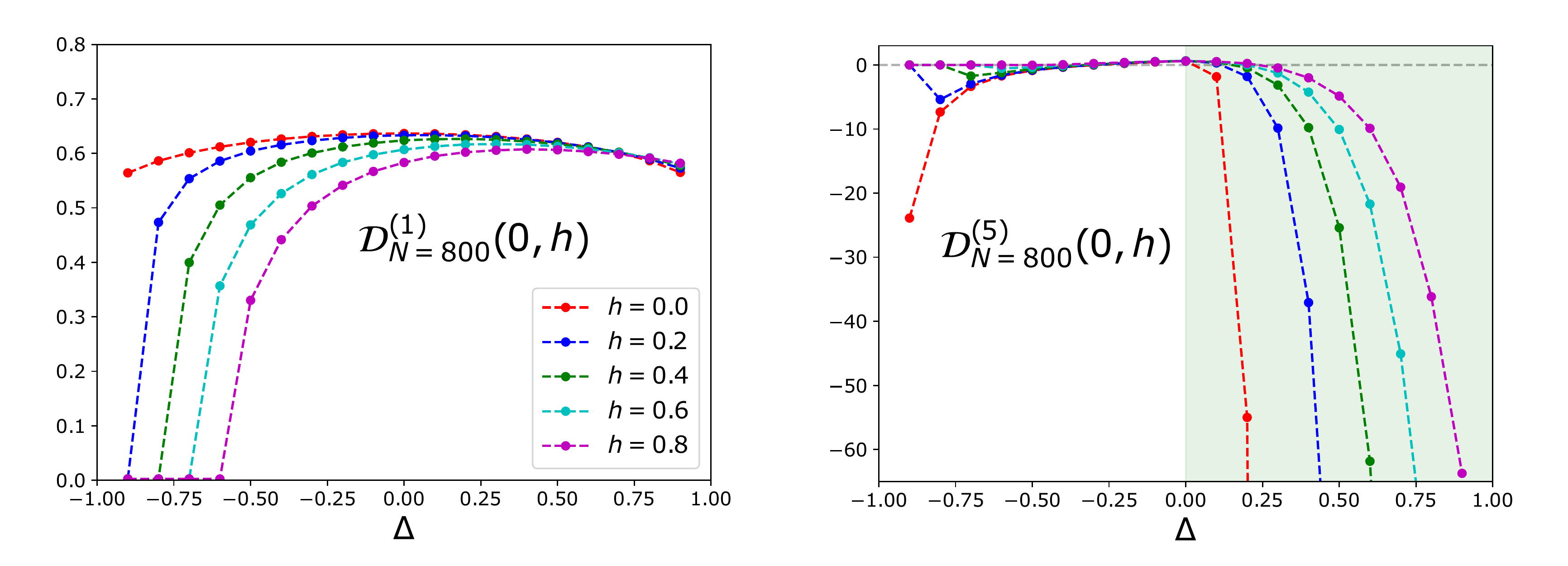}
  \label{D1_h}
  \vspace{-0.25cm}
  \caption{Numerical results for $\mathcal{D}_{N=800}^{(n)}(0,h)$ $(n=1,5)$.
  All the vertical axes are scaled with $J$. Green regions are the divergent regions of NLDWs without a magnetic field, which are determined by $4\gamma/(\pi-\gamma)<n-1$.
  }
\end{figure*}
\begin{figure*}[htbp]
  \centering
  \includegraphics[width=0.865\hsize]{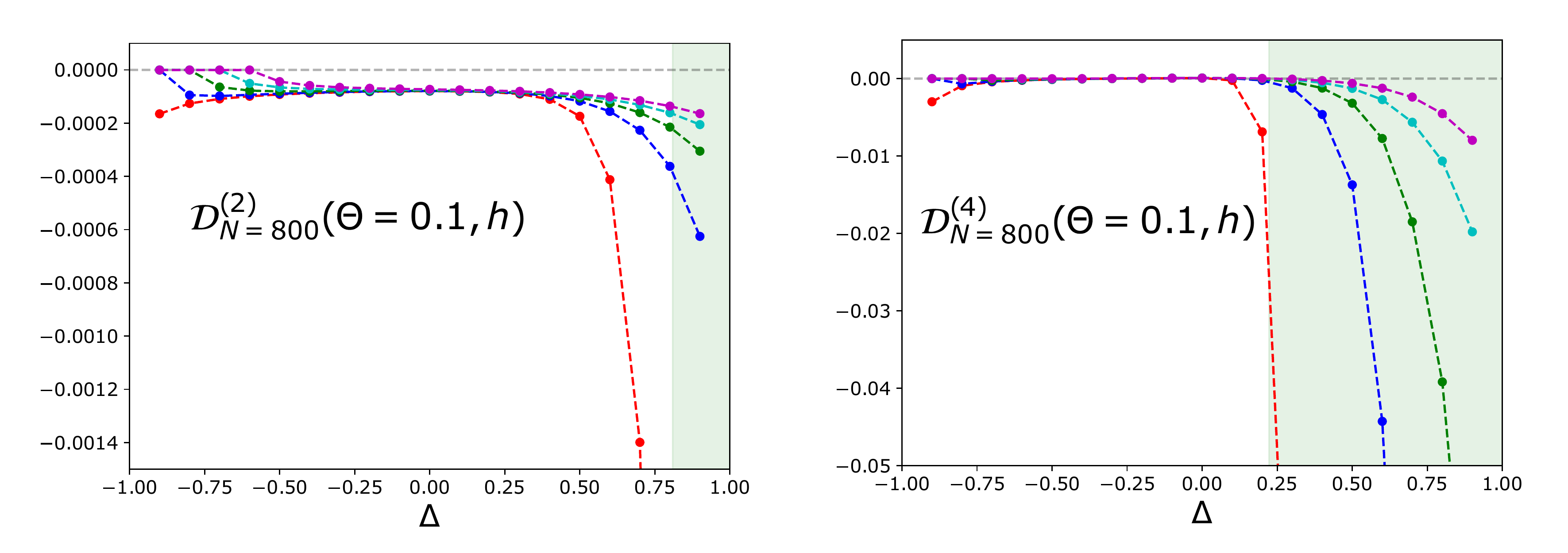}
  \label{D2_h}
  \vspace{-0.25cm}
  \caption{Numerical results for $\mathcal{D}_{N=800}^{(n)}(\Theta=0.1,h)$ $(n=2,4)$.
  All the vertical axes are scaled with $J$. Green regions are the divergent regions of NLDWs without a magnetic field, which are determined by $4\gamma/(\pi-\gamma)<n-1$.}
\end{figure*}

\end{widetext}
\end{document}